%% file: main.tex
\documentclass{article}

\usepackage{microtype}
\usepackage{graphicx}
\usepackage{subfigure}
\usepackage{booktabs} % for professional tables

\usepackage{hyperref}

\usepackage[accepted]{icml2024}

\usepackage{amsmath}
\usepackage{amssymb}
\usepackage{mathtools}
\usepackage{amsthm}

\usepackage[capitalize,noabbrev]{cleveref}

\theoremstyle{plain}

\theoremstyle{definition}

\theoremstyle{remark}

\usepackage[textsize=tiny]{todonotes}

\icmltitlerunning{Full-Atom Peptide Design based on Multi-modal Flow Matching}

\begin{document}

\twocolumn[
\icmltitle{Full-Atom Peptide Design based on Multi-modal Flow Matching}



\icmlsetsymbol{equal}{*}

\begin{icmlauthorlist}
\icmlauthor{Jiahan Li}{equal,yyy,air}
\icmlauthor{Chaoran Cheng}{equal,uiuc}
\icmlauthor{Zuofan Wu}{yyy}
\icmlauthor{Ruihan Guo}{yyy}
\icmlauthor{Shitong Luo}{yyy}
\icmlauthor{Zhizhou Ren}{yyy}
\icmlauthor{Jian Peng}{yyy}
\icmlauthor{Jianzhu Ma}{yyy,air}
\end{icmlauthorlist}

\icmlaffiliation{yyy}{Helixon Research}
\icmlaffiliation{air}{Institute for AI Industry Research, Tsinghua University}
\icmlaffiliation{uiuc}{Department of Computer Science, University of Illinois Urbana-Champaign}

\icmlcorrespondingauthor{Jiahan Li}{lijiahanypc@pku.edu.cn}
\icmlcorrespondingauthor{Jianzhu Ma}{majianzhu@tsinghua.edu.cn}

\icmlkeywords{Machine Learning, ICML}

\vskip 0.3in
]



\printAffiliationsAndNotice{\icmlEqualContribution} % otherwise use the standard text.

\input{Sections/0_abstract}

\input{Sections/1_introduction}

\input{Sections/2_related_work}

\input{Sections/3_methods}

\input{Sections/4_experiment}

\input{Sections/5_conclusion}

\bibliography{ref}
\bibliographystyle{icml2024}

\input{Sections/6_supplementray}

\end{document}

%% file: Sections/0_abstract.tex
\begin{abstract}
Peptides, short chains of amino acid residues, play a vital role in numerous biological processes by interacting with other target molecules, offering substantial potential in drug discovery.
In this work, we present \emph{PepFlow}, the first multi-modal deep generative model grounded in the flow-matching framework for the design of full-atom peptides that target specific protein receptors.
Drawing inspiration from the crucial roles of residue backbone orientations and side-chain dynamics in protein-peptide interactions, we characterize the peptide structure using rigid backbone frames within the $\mathrm{SE}(3)$ manifold and side-chain angles on high-dimensional tori. Furthermore, we represent discrete residue types in the peptide sequence as categorical distributions on the probability simplex.
By learning the joint distributions of each modality using derived flows and vector fields on corresponding manifolds, our method excels in the fine-grained design of full-atom peptides.
Harnessing the multi-modal paradigm, our approach adeptly tackles various tasks such as fix-backbone sequence design and side-chain packing through partial sampling.
Through meticulously crafted experiments, we demonstrate that \emph{PepFlow} exhibits superior performance in comprehensive benchmarks, highlighting its significant potential in computational peptide design and analysis. 
\end{abstract}

%% file: Sections/1_introduction.tex
\section{Introduction}

Peptides, comprising approximately $3$ to $20$ amino-acid residues, are single-chain proteins \cite{bodanszky1988peptide}. By binding to other molecules, especially target proteins (receptors), peptides serve as integral players in diverse biological processes, such as cellular signaling, enzymatic catalysis, and immune responses \cite{petsalaki2008peptide,kaspar2013future}. Therapeutic peptides that bind to diseases-associated proteins are gaining recognition as promising drug candidates due to their strong affinity, low toxicity, and easy delivery \cite{craik2013future,fosgerau2015peptide,muttenthaler2021trends,wang2022therapeutic}. Traditional discovery methods, such as mutagenesis and immunization-based library construction, face limitations due to the vast design space of peptides \cite{lam1997mini,vlieghe2010synthetic,fosgerau2015peptide}. To break the experimental constraints, there is a growing demand for computational methods facilitating \textit{in silico} peptide design and analysis \cite{bhardwaj2016accurate,cao2022design,xie2023helixgan,manshour2023integrating,bryant2023peptide,bhat2023novo}.

\vspace{-1em}
\begin{figure}[!htbp]
    \begin{minipage}{0.5\linewidth}
        \centering     
        \includegraphics[width=\linewidth]{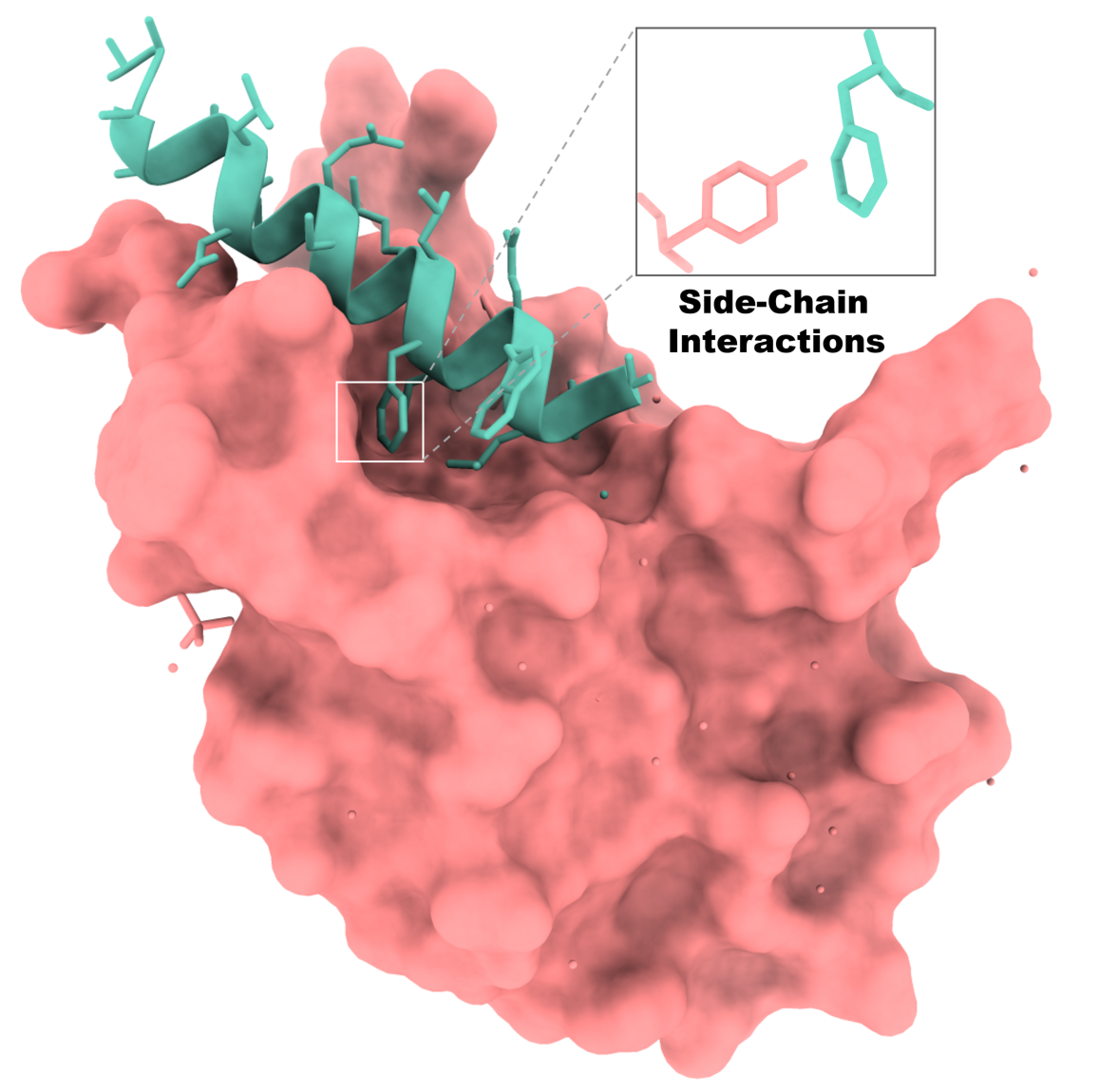}
    \end{minipage}\hfill
    \begin{minipage}{0.5\linewidth}
        \centering
        \includegraphics[width=\linewidth]{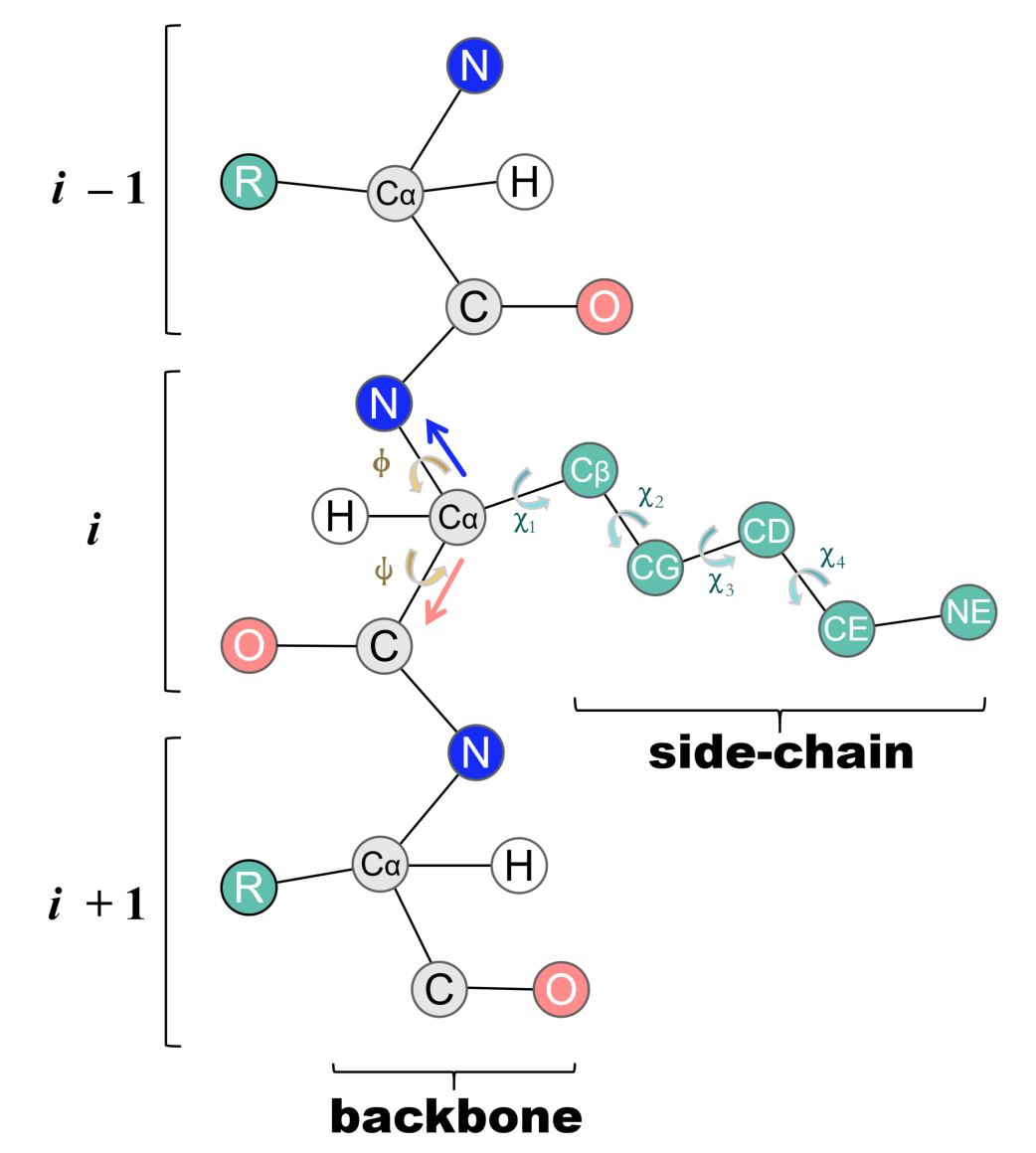}
    \end{minipage}
    \vspace{-1em}
    \caption{\textbf{Left}. A peptide binds to its target protein receptor, highlighting the pivotal role of backbone orientations and side-chain interactions among key residues. \textbf{Right}. Every protein residue consists of backbone atoms and side-chain atoms. The backbone atoms establish a rigid frame, whereas the side-chain atoms contribute to flexible side-chain angles.}
    \label{fig:protein}
\end{figure}
\vspace{-1em}

% 2. Computational protein design, backbone, diffusion
Recently, deep generative models, particularly diffusion probabilistic models \cite{sohl2015deep,ho2020denoising,song2019generative,song2020score}, have shown considerable promise in \textit{de novo} protein design \cite{huang2016coming}. These models mainly focus on generating protein backbones, represented as $N$ rigid frames in the $\mathrm{SE}(3)$ manifold \cite{trippe2022diffusion,anand2022protein,luo2022antigen,wu2022protein,ingraham2023illuminating,yim2023se}. The current state-of-the-art method, RFDiffusion \cite{watson2022broadly} excels in designing a diverse array of functional proteins with an enhanced success rate validated through experiments \cite{roel2023single}.

% 3. however, problems: 1. target information, binding site 2. side-chain full atom 3. seq-struct, second struct 4. diffusion?
Though achieving remarkable success in protein backbone design, current generative models still encounter challenges in generating peptide binders conditioned on a specific target protein \cite{bennett2023improving}. Unlike unconditional generation, generating binding peptides necessitates explicit \emph{conditioning on binding pockets}, as the bound-state structures of peptides in protein-peptide complexes partially depend on their targets \cite{duffaud1985structure,dagliyan2011structural} and peptides contact accurate binding sites for biological functions. Furthermore, as shown in Figure \ref{fig:protein}, in protein-peptide interactions \cite{stanfield1995protein,jacobson2002role}, the focus should extend beyond the positions and orientations of the backbones to encompass the dynamics of side-chain angles, where residues interact with each other through non-covalent forces formed by side-chain groups. Consequently, peptide generation should \emph{consider full-atom structures} rather than solely concentrating on modeling the four backbone heavy atoms. Also, since the structure of the functional peptide is mainly determined by its sequence \cite{whisstock2003prediction}, it is essential to simultaneously \emph{consider sequence and structure} during generation to enhance consistency between them. 

% Lastly, current diffusion models for proteins suffers from long sampling time and unstable convergence.

% Peptides contact the accurate binding site on target for their biological functions. 
% And generated peptides should bind to the accurate binding site on the target protein to effectively fulfill their biological functions.

To address the challenges mentioned above, we introduce \emph{PepFlow}, a multi-modal deep generative model built upon the Conditional Flow Matching (CFM) framework \cite{lipman2022flow}. CFM learns the continuous normalizing flow \cite{chen2018neural} by regressing the vector field that transforms prior distributions to target distributions. It has demonstrated competitive generation performance compared to the diffusion framework and is handy to adopt on non-Euclidean manifolds \cite{chen2023riemannian}. We further extend CFM to modalities related to full-atom proteins. In our framework, each residue in the peptide is represented as a rigid frame in $\mathrm{SE}(3)$ for backbone, a point on a hypertorus corresponding to side-chain angles, and an element on the probability simplex indicating the discrete type. We derive analytical flows for each modality and model the joint distribution of the full-atom peptide structure and sequence conditioned on the target protein. Subsequently, we can design full-atom peptides by simultaneously transforming from the prior distribution to the learned distribution on each modality. Our method extends its applicability to other tasks including fix-backbone sequence design and side-chain packing, which is achieved through partial sampling for desired modalities. Notably, there is no current comprehensive benchmark for evaluating generative models in peptide design, we further introduce a new dataset with in-depth metrics to quantify the qualities of generated peptides.

In summary, our key contributions include:
% 1) We extend flow matching to protein-related modalities, especially toric space and probability simplex.
\begin{itemize}
    \item We introduce \emph{PepFlow}, the first multi-modal generative model for designing full-atom protein structures and sequences
    \item We pioneer the resolution of challenges in target-specific peptide design and establish comprehensive benchmarks, including newly cleaned dataset and novel \textit{in silico} metrics to evaluate generated peptides.
    \item Our method showcases superior performance and great scalability across a spectrum of peptide design and analysis tasks encompassing sequence-structure co-design, fix-backbone sequence design, and side-chain packing.
\end{itemize}

%% file: Sections/2_related_work.tex
\section{Related Work}

\textbf{Diffusion-based and Flow-based Generative Models} Trained on the denoising score matching objective \cite{vincent2011connection}, diffusion models refine samples from prior Gaussian distributions into gradually meaningful outputs \cite{sohl2015deep, ho2020denoising, song2019generative, song2020score}. Diffusion models are applicable in diverse modalities, e.g. images \cite{ho2020denoising, zhang2023hive}, texts \cite{austin2021structured,hoogeboom2021argmax} and molecules \cite{hoogeboom2022equivariant, xu2022geodiff, jing2022torsional}. However, they rely on stochastic estimation which leads to suboptimal probability paths and longer sample steps \cite{song2020denoising, lu2022dpm}.

Another direction considers ODE-based continuous normalizing flows as an alternative to diffusion models \cite{chen2018neural}. Conditional Flow Matching (CFM) \cite{lipman2022flow, liu2022flow, albergo2022building} directly learns the ODE that traces the probability path from the prior distribution to the target, regressing the pushing-forward vector field conditioned on individual data points. Additionally, Riemannian Flow Matching \cite{chen2023riemannian} extends CFM to general manifolds without the requirement for expensive simulations \cite{ben2022matching,de2022riemannian,huang2022riemannian}. In our work, we use the flow matching framework to conditionally model full-atom structures and sequences of peptide binders. Though previous work existed in applying flow matching models in molecular generation \cite{song2023equivariant, bose2023se, yim2023fast, yim2024improved}, they mainly focused on specific modalities or unconditional generation.

% summarize our work? other flow-based work on protein?
\textbf{\textit{De novo} Protein Design with Generative Models} Generative models have demonstrated promising performance in the design of protein-related applications including enzyme active sites \cite{yeh2023novo,dauparas2023atomic,zhang2023full} and motif scaffolds \cite{wang2021deep,trippe2022diffusion,yim2024improved}. These methods can be categorized into three main schemes: \emph{sequence design}, \emph{structure design}, and \emph{sequence-structure co-design}. In sequence design, protein sequences are crafted using oracle-based directed evolution \cite{jain2022biological,ren2022proximal,khan2022antbo,stanton2022accelerating}, protein language models \cite{madani2020progen,verkuil2022language,nijkamp2023progen2}, and discrete diffusion models \cite{alamdari2023protein,frey2023protein,gruver2024protein,yi2024graph}.Alternatively, sequences are sampled based on protein backbone structures, known as \emph{fix-backbone sequence design} \cite{ingraham2019generative,jing2020learning,hsu2022learning,li2022orientation,gao2022pifold}. Recognizing the important role of protein 3D structures, another approach involves directly generating protein backbone structures \cite{trippe2022diffusion,anand2022protein,luo2022antigen,wu2022protein,ingraham2023illuminating}. These backbones are then fed into fix-backbone design models to predict the corresponding sequences, e.g. ProteinMPNN \cite{dauparas2022robust}. Sequence-structure co-design methods jointly sample sequence-structure pairs conditioned on provided information and find widespread usage in designing antibodies \cite{jin2021iterative,luo2022antigen,kong2022conditional}. Nevertheless, few methods focus on the interplay of side-chain interactions in the conditional generation of proteins or concurrently producing protein structures and sequences in full atomic detail \cite{martinkus2023abdiffuser,kong2023end,krishna2023generalized}.

%% file: Sections/3_methods.tex
\section{Methods}

\subsection{Preliminary}
A protein is a biomolecule comprised of several amino acid residues, each characterized by its type, backbone frame, and side-chain angles \cite{fisher2001lehninger}, as illustrated in Figure \ref{fig:protein}. The type of the $i$-th residue $a^i \in \{1...20\}$ is determined by its side-chain R group. The rigid frame is constructed by using the coordinates of four backbone heavy atoms N-C$_{\alpha}$-C-O, with C$_{\alpha}$ located at the origin. Thus, a residue frame is parameterized by a position vector $\mathbf{x}^i \in \mathbb{R}^3$ and a rotation matrix $R^i \in \mathrm{SO}(3)$ \cite{jumper2021highly}. The side-chain conformation exhibits flexibility compared to the rigid backbone and can be represented as up to four torsion angles corresponding to rotatable bonds between side-chain atoms $\chi_i \in [0,2\pi)^4$. We further consider the rotatable backbone torsion angle $\psi^i \in [0,2\pi)$ which governs the position of the O atom. Consequently, a protein with $n$ residues can be sufficiently and succinctly parameterized as $\{(a^i,R^i,\mathbf{x}^i,\chi^i)\}_{i=1}^n$, where $\chi^i[0]=\psi^i$ and $\chi^i \in [0,2\pi)^{5}$.

In this work, we focus on designing peptides based on their target proteins. Formally, given a $n$-residue peptide $C^{\text{pep}}=\{(a^j,R^j,\mathbf{x}^j,\mathbf{\chi}^{j})\}_{j=1}^n$ and its $m$-residue target protein (receptor) $C^{\text{rec}} = \{(a^i,R^i,\mathbf{x}^i,\mathbf{\chi}^i)\}_{i=1}^m$, we aim to model the conditional joint distribution $p(C^{\text{pep}} | C^{\text{rec}})$. 

%We describe our multi-modal flow matching method for modeling this conditional joint distribution in the following sections.

\subsection{Multi-modal Flow Matching}

\begin{figure*}[!htbp]
    \centering
    \includegraphics[width=0.9\linewidth]{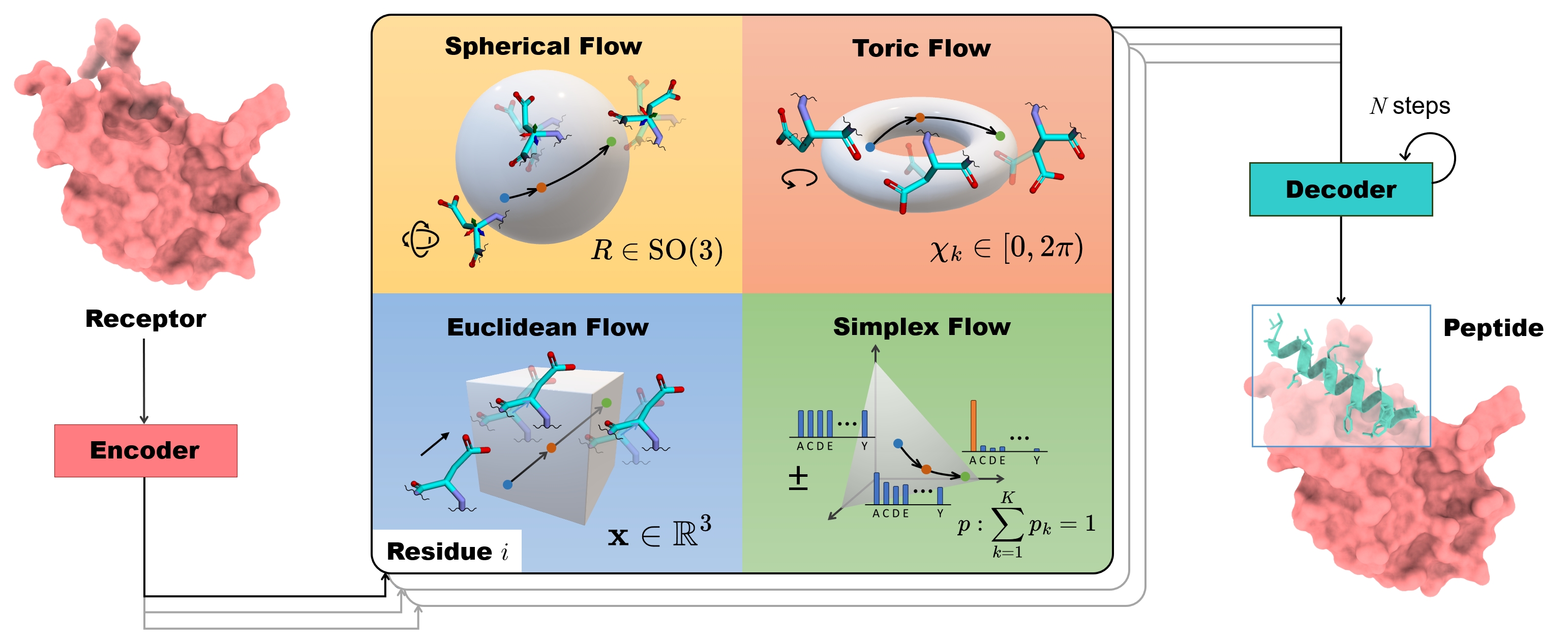}
    \vspace{-1em}
    \caption{Illustration of PepFlow Architecture. The encoder encodes the receptor as the context for peptide generation. Flows for four different modalities are then constructed: spherical for the orientation $R$, Euclidean for the translation $\mathbf{x}$, toric for the torsion angles $\chi_k$, and categorical for the type distribution $p$. The multi-modal flow matching decoder finally recovers the full-atom peptide structure and sequence iteratively using the Euler method.
    }
    \label{fig:model}
    % \vspace{-1em}
\end{figure*}

% brief flow matching
The conditional flow matching framework \cite{lipman2022flow} provides a simple yet powerful way to learn a probability flow $\psi$ that pushes the source distribution $p_0$ to the target distribution $p_1$ of the data points $\mathbf{x} \in \mathbb{R}^d$. The time-dependent flow $\psi_t:[0,1]\times \mathbb{R}^d \rightarrow \mathbb{R}^d$ and the associated vector field $u_t:[0,1]\times \mathbb{R}^d \rightarrow \mathbb{R}^d$ can be defined via the Ordinary Differential Equation (ODE) $\frac{d}{dt}\mathbf{x}_t = u_t(\mathbf{x}_t)$ where $\mathbf{x}_t = \psi_t(\mathbf{x_0})$. A time-dependent network $v_\theta(\mathbf{x}_t,t)$ can be used to directly regress the defined vector field, known as the Flow Matching objective (FM). However, the FM objective is intractable in practice, as we have no access to the closed-form vector field $u_t$. Nonetheless, when conditioning the time-dependent vector field and probability flow on the specific sample $\mathbf{x}_1 \sim p_1(\mathbf{x}_1)$ and the prior sample $\mathbf{x}_0 \sim p_0(\mathbf{x}_0)$, we can model $\mathbf{x}_t=\psi_t(\mathbf{x}_0|\mathbf{x}_1)$ and $u_t(\mathbf{x}_t|\mathbf{x}_0,\mathbf{x}_1)=\frac{d}{dt}\mathbf{x}_t$ with the tractable Conditional Flow Matching (CFM) objective:
\begin{equation}
    \mathcal{L}(\theta)=\mathbb{E}_{t,p_1(\mathbf{x}_1),p_0(\mathbf{x}_0)}\|v_\theta(\mathbf{x}_t,t)-u_t(\mathbf{x}_t|\mathbf{x}_1,\mathbf{x}_0)\|^2
\end{equation}
where $t \sim \mathcal{U}(0,1)$. It has been proved that the FM and CFM objectives have identical gradients with respect to the parameters $\theta$, and we can integrate the learned vector field through time for sampling. \citet{chen2023riemannian} further extended CFM to general geometries.

% extend to multimodal
% single residue x_i for simplicity in the following part
%As there are different types of modalities in the joint distribution of the peptide, we empirically factorize the joint distribution into four types of distributions:
We employ the conditional flow matching framework to learn the conditional distribution of an $n$-residue peptide based on its $m$-residue target protein $p(C^{\text{pep}} | C^{\text{rec}})$. We empirically decompose the joint probability into the product of probabilities of four basic elements that can describe the sequence and structure:
\begin{equation}
    \begin{aligned}
        p(C^{\text{pep}} | C^{\text{rec}}) &\propto p(\{a^j\}_{j=1}^n | C^{\text{rec}})p(\{R^j\}_{j=1}^n | C^{\text{rec}}) \\
                             &\quad \cdot p(\{\mathbf{x}^j\}_{j=1}^n | C^{\text{rec}})p(\{\chi^j\}_{j=1}^n | C^{\text{rec}})
    \end{aligned}
\end{equation}
We elaborate on the construction of different probability flows on the residue's position $p(x^j | C^{\text{rec}})$, orientation $p(R^j | C^{\text{rec}})$, torsion angles $p(\mathbf{\chi}^j | C^{\text{rec}})$, and type $p(a^j | C^{\text{rec}})$ as follows. For simplicity, we initially focus on the single $j$-th residue in the peptide.

\paragraph{Euclidean CFM for Position} 
We first adopt Euclidean CFM for the position vector $\mathbf{x}^j \in \mathbb{R}^3$. As a common practice, we choose the isotropic Gaussian $\mathcal{N}(0,I_3)$ as the prior, and our target distribution is $p(\mathbf{x}^j | C^{\text{rec}})$. The conditional flow is defined as linear interpolation connecting sampled noise $\mathbf{x}_0^j \sim \mathcal{N}(0,I_3)$ and the data point $\mathbf{x}_1^j \sim p(\mathbf{x}^j | C^{\text{rec}})$. The linear interpolation favors a straight trajectory, and such a property contributes to the efficiency of training and sampling, as it is the shortest path between two points in Euclidean space \cite{liu2022flow}. The conditional vector field $u_t^{\text{pos}}$ is derived by taking the time derivative of the linear flow $\psi_t^{\text{pos}}$:
\begin{align}
\psi_t^{\text{pos}}(\mathbf{x}_0^j|\mathbf{x}_1^j)&=t\mathbf{x}_1^j+(1-t)\mathbf{x}_0^j \\
    u_t^{\text{pos}}(\mathbf{x}_t^j|\mathbf{x}_1^j,\mathbf{x}_0^j)&=\mathbf{x}_1^j-\mathbf{x}_0^j=\frac{\mathbf{x}_1^j-\mathbf{x}_t^j}{1-t}
\end{align}
We use a time-dependent translation-invariant neural network $v^{\text{pos}}$ to predict the conditional vector field based on the current interpolant $\mathbf{x}_t$ and the timestep $t$. The CFM objective of the $j$ th residue is formulated as:
\begin{equation}
    \mathcal{L}^{\text{pos}}_j=\mathbb{E}_{t,p(\mathbf{x}_1^j),p(\mathbf{x}_0^j)}\left\|v^{\text{pos}}(\mathbf{x}_t^j,t,C^{\text{rec}})-(\mathbf{x}_1^j-\mathbf{x}_0^j)\right\|_2^2 \label{eqn:loss_pos}
\end{equation}
During generation, we first sample from the prior $\mathbf{x}^j_0 \sim \mathcal{N}(0,I_3)$ and solve the probability flow with the learned predictor $v^\text{pos}$ using the $N$-step forward Euler method to get the position of residue $j$ with $t=\{0,...,\frac{N-1}{N}\}$:
\begin{equation}
    \mathbf{x}_{t+\frac{1}{N}}^j = \mathbf{x}_t^j+\frac{1}{N}v^{\text{pos}}(\mathbf{x}_t^j,t,C^{\text{rec}})
\end{equation}

\paragraph{Spherical CFM for Orientation} 
The orientation of the residue $j$ can be represented as a rotation matrix $R^j\in\mathrm{SO}(3)$ concerning the global frame. The 3D rotation group $\mathrm{SO}(3)$ is a smooth Riemannian manifold and its tangent $\mathfrak{so}(3)$ is a Lie group containing skew-symmetric matrices. An element in $\mathfrak{so}(3)$ can also be interpreted as an infinitesimal rotation around a certain axis and characterized as a rotation vector in $\mathbb{R}^3$ \cite{blanco2021tutorial}. We choose the uniform distribution over $\mathrm{SO}(3)$ as our prior distribution. Just as flow matching in Euclidean space uses the shortest distance between two points, we can envision establishing flows by following the corresponding \emph{geodesics} in the context of $\mathrm{SO}(3)$ \cite{lee2018introduction}. Geodesics on $\mathrm{SO}(3)$ represent the paths of the minimum rotational distance between two orientations and provide a natural way to interpolate and evolve orientations in a manner that respects the underlying geometry of the rotation manifold \cite{bose2023se,yim2023fast}. The conditional flow $\psi^{\text{ori}}$ and vector field $u_t^{\text{ori}}$ are established by the geodesic interpolation between $R^j_0 \sim U(\mathrm{SO}(3))$ and $R_1^j \in p(R^j | C^{\text{rec}})$ with the geodesic distance decreasing linearly by time:
\begin{align}
    \psi^{\text{ori}}_t(R^j_0|R^j_1)&=\mathtt{exp}_{R_0^j}(t\mathtt{log}_{R_0^j}(R_1^j)) \\
    u_t^{\text{ori}}(R_t^j|R_0^j,R_1^j)&=\frac{\mathtt{log}_{R_t^j}R_1^j}{1-t}
\end{align}
where $\mathtt{exp}$ and $\mathtt{log}$ are the exponential and logarithm maps on $\mathrm{SO}(3)$ that can be computed efficiently using Rodrigues' formula (see Appendix \ref{sec:manifold}). A rotation-equivariant neural network $v^{\text{ori}}$ is applied to predict the vector field, represented as rotation vectors. The CFM objective on $\mathrm{SO}(3)$ is formulated as:
\begin{equation}
    \mathcal{L}_{\text{ori}}^j = \mathbb{E}_{t,p(R_1^j),p(R_0^j)}\left\|v^{\text{ori}}(R_t^j,t,C^{\text{rec}})-\frac{\mathtt{log}_{R_t^j}R_1^j}{1-t}\right\|_{\mathrm{SO}(3)}^2
\end{equation}
where the vector field lies on the tangent space $\mathfrak{so}(3)$ of $\mathrm{SO(3)}$ and the norm is induced by the canonical metric on $\mathrm{SO(3)}$. % Here we parameterize the vector field as its axis-angle vector representation in $\mathbb{R}^3$,
During inference, we initiate the process from $R^j_0\sim U(\mathrm{SO}(3))$ and proceed by taking small steps along the geodesic path in $\mathrm{SO}(3)$ over the timestep $t$:
\begin{equation}
    R_{t+\frac{1}{N}}^j = \mathtt{exp}_{R_t^j}\left(\frac{1}{N}v^{\text{ori}}(R_t^j,t,C^{\text{rec}})\right) \label{eqn:loss_ori}
\end{equation}

% some shit on SE3? zero center of mass

\paragraph{Toric CFM for Angles} 
An angle taking values in $[0,2\pi)$ can be represented as a point on the unit circle $\mathbb{S}^1$, and the torsion vector $\chi^j$ consisting $5$ angles lies on the $5$-dimensional flat torus $\mathbb{T}^d=(\mathbb{S}^1)^{5}$ as the Cartesian product of $5$ one-dimensional unit circles. The flat torus $\mathbb{T}^5$ can also be viewed as the quotient space $\mathbb{R}^5/(2\pi\mathbb{Z})^5$ that inherits its Riemannian metric from Euclidean space. Therefore, the exponential and logarithm maps are similar to those in Euclidean space except for the equivalence relation about the periodicity of $2\pi$.
%Here $a,b\in\mathbb{T}^d,u\in \mathbb{R}^d$ and functions are taken element-wise.
We choose the uniform distribution on $[0,2\pi]^5$ as the prior distribution, and the conditional flow between sampled points $\chi^j_0 \sim U([0,2\pi]^5)$ and $\chi^j_1 \sim p(\chi^j|C^{\text{rec}})$ is constructed along the geodesic:
\begin{align}
\text{wrap}(u)&=(u+\pi) \%(2\pi)-\pi \\
    \psi_t^{\text{ang}}(\chi_0^j|\chi_1^j)&=(\chi_0^j+t(\chi_1^j-\chi_0^j))\%2\pi \\
     u_t^{\text{ang}}(\chi_t^j|\chi_0^j,\chi_1^j)&=\text{wrap}\left(\frac{\chi_1^j-\chi_t^j}{1-t}\right)
    % u_t^{\text{ang}}(\chi_t^j|\chi_0^j,\chi_1^j)&=\frac{(\chi_1^j-\chi_0^j)\% 2\pi}{1-t}
\end{align}
A neural network $v^{\text{ang}}$ is applied to predict the vector field that lies on the tangent space of $\mathbb{T}^5$, leading to the CFM objective on torus as:
\begin{equation}
    \mathcal{L}_{\text{ang}}^j = \mathbb{E}_{t,p(\mathbf{\chi}_1^j),p(\mathbf{\chi}_0^j)}\left\|\text{wrap}\left(v^{\text{ang}}(\mathbf{\chi}_t^j,t,C^{\text{rec}})-\frac{\chi_1^j-\chi_0^j}{1-t}\right)\right\|^2 
    % \mathcal{L}_{\text{ang}}^j = \mathbb{E}_{t,p(\mathbf{\chi}_1^j),p(\mathbf{\chi}_0^j)}\left\|\left(v^{\text{ang}}(\mathbf{\chi}_t^j,t,C^{\text{rec}})-\frac{\chi_1^j-\chi_0^j}{1-t}\right)\%2\pi\right\|^2 
    \label{eqn:loss_chi}
\end{equation}
Instead of using the standard distance of the Euclidean space, we use the flat metric on $\mathbb{T}^d$ for comparing the predicted and ground truth vector field, which enhances the network's awareness of angular periodicity on the toric geometry. During inference, we take Euler steps from prior sample points $\chi^j \sim U([0,2\pi]^5)$ along the geodesic path in $\mathbb{T}^5$:
\begin{equation}
    \chi^j_{t+\frac{1}{N}}=\left(\chi^j_t+\frac{1}{N}v^{\text{ang}}(\mathbf{\chi}_t^j,t,C^{\text{rec}})\right)\%2\pi\label{eqn:logit}
\end{equation}

\paragraph{Simplex CFM for Type} 
The above flow-matching methods apply to values in continuous spaces. The residue type $a^j \in \{1...20\}$, however, admits a discrete categorical value. To map the discrete residue type $a^j$ to a continuous space, we adopt a soft one-hot encoding operation $\mathtt{logit}(a^j)=\mathbf{s}^j \in \mathbb{R}^{20}$ with a constant value $K>0$, and the $i$-th value in $\mathbf{s}^j$ is
\begin{equation}
\mathbf{s}^j[i]=
\begin{cases}
    K, &i=a^j\\
    -K,&\text{otherwise}
\end{cases}
\end{equation}
$\mathbf{s}^j$ can be understood as the logits of the probabilities \cite{han-etal-2023-ssd}, and $\mathtt{softmax}(\mathbf{s}^j)$ becomes a normalized probability distribution with the $j$-th term close to $1$ and other terms close to $0$. This representation promotes the underlying categorical distribution with a probability mass centered on the correct residue type $a^j$. In other word, $\mathtt{softmax}(\mathbf{s}^j)$ is a point on the $20$-category probability simplex $\Delta^{19}$. The $d$-categorical probability simplex $\Delta^{d-1}:=\{\mathbf{x}\in \mathbb{R}^d:0\leq \mathbf{x}[i]\leq 1,\sum_{i=1}^d \mathbf{x}[i]=1\}$ is a smooth manifold in $\mathbb{R}^d$ \cite{wang2013projection,richemond2022categorical,floto2023diffusion}, where each point in $\Delta^{d-1}$ represents a categorical distribution over the $d$ classes. Though we can directly perform flow matching on $\Delta^{d-1}$ \cite{li2018geometry}, we choose to construct conditional flow on the logit space in $\mathbb{R}^d$. We choose our prior distribution of logit as $\mathcal{N}(0,K^2 I)$ such that the prior distribution on simplex becomes the logistic-normal distribution by construct \cite{atchison1980logistic}. The conditional flow $\psi_t^{\text{type}}$ and vector field on the logit space is defined as:
\begin{align}
\psi_t^{\text{type}}(\mathbf{s}_0^j|\mathbf{s}_1^j)&=t\mathbf{s}_1^j+(1-t)\mathbf{s}_0^j\\
    u_t^{\text{type}}(\mathbf{s}_t^j|\mathbf{s}_1^j,\mathbf{s}_0^j)&=\mathbf{s}_1^j-\mathbf{s}_0^j=\frac{\mathbf{s}_1^j-\mathbf{s}_t^j}{1-t}
\end{align}
The linear interpolations between $\mathbf{s}^j_1$ and $\mathbf{s}^j_0$ induce a geometric mean of the prior logit-normal distribution and target distribution $p(a_j|C^{\text{rec}})$. This induces a time-dependent interpolant between points on the probability simplex, capturing the evolving relationship between the prior and target distributions. Similarly, a neural network $v^{\text{type}}$ is applied to predict the vector field on the logit space, and the CFM objective is:
\begin{equation}
    \mathcal{L}^{\text{type}}_j=\mathbb{E}_{t,p(\mathbf{s}_1^j),p(\mathbf{s}_0^j)}\left\|v^{\text{type}}(\mathbf{s}_t^j,t,C^{\text{rec}})-(\mathbf{s}_1^j-\mathbf{s}_0^j)\right\|_2^2 \label{eqn:loss_cls}
\end{equation}

During inference, we perform Euler steps to solve the probability flow on the logit space, residue types can be sampled from the corresponding probability vector on the simplex:
\begin{align}
    \mathbf{s}_{t+\frac{1}{N}}^j &= s_t^j+\frac{1}{N}v^{\text{type}}(\mathbf{s}_t^j,t,C^{\text{rec}}) \\
    a^j_{t+1/N}&\sim\mathtt{softmax}(\mathbf{s}_{t+\frac{1}{N}}^j)
\end{align}
To improve generation consistency between the logit and simplex space, we additionally map the predicted discrete residue back to the logit space during each iteration as $\mathbf{s}_{t+1/N}^j=\mathtt{logit}(a^j_{t+\frac{1}{N}})$.

Combining all modalities, we obtain the final multi-modal conditional flow matching objective for residue $j$ as the weighted sum of different conditional flow matching objectives:

\vspace{-1.5em}
\begin{equation}
    \mathcal{L}^j_\text{cfm}=\mathbb{E}_t(\lambda^{\text{pos}}\mathcal{L}^{\text{pos}}_j
    +\lambda^\text{ori}\mathcal{L}^\text{ori}_j
    +\lambda^\text{ang}\mathcal{L}^\text{ang}_j
    +\lambda^\text{type}\mathcal{L}^\text{type}_j)\label{eqn:vf_loss}
\end{equation} 
\vspace{-1.5em}

After discussing the multi-modal flow for modeling the factorized distribution of position, orientation, residue type, and side-chain torsion angles for a single residue $j$, we further extend our method for modeling the joint distribution $p(C^{\text{pep}}|C^{\text{rec}})$ in the following subsection.

\subsection{PepFlow Architecture}

\paragraph{Network Parametrization}
As the joint distribution of the peptide is conditioned on its target protein receptor, we employ a geometric equivariant encoder to capture the context information of the target protein. Conversely, the above flow matching model can be viewed as a decoder for regressing the vector fields of the generated peptide, as depicted in Figure \ref{fig:model}. In the model pipeline, the encoder $\mathrm{Enc}$ takes the sequence and structure of the target protein $C^{\text{rec}}$, producing the hidden residue representations $\mathbf{h}$ and the residue-pair embedding $\mathbf{z}$. Subsequently, we sample a specific timestep $t \sim U(0,1)$ to construct the multi-modal flows where time-dependent vector fields are learned simultaneously for each modality of the peptide $C^{\text{pep}}$. The time-dependent decoder network $\mathrm{Dec}$, mainly based on the Invariant Point Attention scheme \cite{jumper2021highly}, takes the timestep $t$, the interpolant state of the peptide $C^{\text{pep}}_t=\{(\mathbf{s}^j_t,R^j_t,\mathbf{x}^j_t,\chi^j_t)\}_{j=1}^n$, and the residue and pair embeddings as input. 
Rather than directly regressing the vector fields, the decoder first recovers the original peptide $\overline{C}^{\text{pep}}_1=\{(\overline{\mathbf{s}}^j_1,\overline{R}^j_1,\overline{\mathbf{x}}^j_1,\overline{\chi}^j_1)\}_{j=1}^n$ which allows for better training efficiency and the use of auxiliary loss. In this way, the CFM objectives for different modalities are reparametrized as the distance between the vector fields derived from ground truth elements and those derived from predicted elements (see Appendix \ref{sec:reparam}). The overall training objective is the sum of reparameterized CFM objectives, considering the expectation over each timestep and each residue in the peptide.
\begin{equation}
    \mathcal{L}=\mathbb{E}_{t}\left[\frac{1}{n}\sum_j (\mathcal{L}^j_\text{cfm} + \lambda^{\text{aux}} (\mathcal{L}^j_\text{bb} + \mathcal{L}^j_\text{tor}))\right]
    \label{eqn:final_loss}
\end{equation}
Here we also use the backbone position loss $\mathcal{L}_{\text{bb}}$ and the torsion angle loss $\mathcal{L}_{\text{tor}}$ as auxiliary structure losses to improve the generation quality, incorporating information from different modalities (see Appendix \ref{sec:aux_loss}). The training process is outlined in Algorithm \ref{alg:train}.

\begin{algorithm}
    % \vspace{-1em}
    \caption{Training Multi-Modal PepFlow}\label{alg:train}
    \begin{algorithmic}[1]
        \WHILE {not converged}
            \STATE Sample protein-peptide pair $C^{\text{rec}},C^{\text{pep}}$ from dataset
            \STATE Encode target $\mathbf{h,z}=\mathrm{Enc}(C^{\text{rec}})$
            \STATE Sample prior state $C^{\text{pep}}_0=\{(a^j_0,R^j_0,\textbf{x}^j_0,\chi^j_0)\}_{j=1}^{n}$
            \STATE Sample $t\sim U(0,1)$
            \STATE Decode predicted peptide $\overline{C}^{\text{pep}}=\mathrm{Dec}(C^{\text{pep}}_{t},t,\mathbf{h,z})$
            \STATE Calculate the vector fields and the loss according to Eq.\eqref{eqn:final_loss}
            \STATE Update the parameters of $\mathrm{Enc}$ and $\mathrm{Dec}$
        \ENDWHILE
    \end{algorithmic}
\end{algorithm}
\vspace{-1em}

\paragraph{Sampling Process}
% sample, conditional other modalities and residues
% partial sample
The sampling algorithm is outlined in Algorithm \ref{alg:sample}. During sampling, the target protein is encoded only once and is fed into each step of decoding. Initially, a prior state of the peptide is sampled. Subsequently, for each timestep of the sampling, the decoder predicts the recovered peptide state, and vector fields are calculated for each modality of the current state. The current state of the peptide is then updated using the Euler method following the derived vector fields, and the updated peptide is considered as the input of the decoder in the subsequent iteration. In the final step, we reconstruct the full-atom peptide using local coordinate transformations of the backbone and side-chain rigid groups \cite{jumper2021highly}. 

Noticeably, for a specific residue, the generation of a particular modality is not only dependent on the update of that modality but also influenced by the states of other modalities and other residues within the peptide. This interdependence highlights the intricate relationship between different modalities and residues, illustrating the complex nature of the joint distribution captured by our designed decoder network.

\begin{figure}[t]
\vspace{-1em}
\begin{algorithm}[H]
    % \vspace{-0.5em}
    \caption{Sampling with Multi-Modal PepFlow}\label{alg:sample}
    \begin{algorithmic}[1]
        % \STATE \textbf{Input:} Target $C^{\text{rec}}$, Length $n$, Steps $N$, Encoder $\mathrm{Enc}$, Decoder $\mathrm{Dec}$
        % \STATE \textbf{Output:} $\overline{C}^{\text{pep}}_1$
        \STATE Encode target $\mathbf{h,z}=\mathrm{Enc}(C^{\text{rec}})$
        \STATE Sample prior state $C^{\text{pep}}_0=\{(a^j_0,R^j_0,\textbf{x}^j_0,\chi^j_0)\}_{j=1}^{n}$
        \FOR {$t\gets 1$ to $N$}
            \STATE Decode predicted peptide $\overline{C}^{\text{pep}}_{\frac{t}{N}}=\mathrm{Dec}(C^{\text{pep}}_{\frac{t-1}{N}},t,\mathbf{h,z})$
            \STATE Calculate vector fields and update the peptide $C^{\text{pep}}_\frac{t}{N}=\mathrm{EulerStep}(\overline{C}^{\text{pep}}_{\frac{t}{N}},C^{\text{pep}}_{\frac{t-1}{N}},\frac{1}{N})$
        \ENDFOR
        \STATE {\textbf{return} $\overline{C}^{\text{pep}}_1$}
    \end{algorithmic}
\end{algorithm}
\vspace{-3.em}
\end{figure}

Furthermore, beyond the joint design of sequences and structures, we can construct partial states tailored for peptide design tasks that specifically emphasize the generation of a particular modality while keeping other modalities fixed. In the context of side-chain packing, we maintain the ground truth sequence and the backbone structure (i.e., types, positions, and orientations), and exclusively sample the torsion angles. Conversely, in fix-backbone sequence design, our focus lies in sampling sequences while holding constant the backbone positions and orientations.
% For flexible docking, the sequences are known and the emphasis is on structural sampling. We evaluate our method with comprehensive benchmarks in the next section.

% network, predict gt, aux loss, SE(3) and symmetric, 

% We have described the flow-matching architecture of the proposed model in detail. However, the generation of the ligand peptide is closely related to the pocket geometry and physicochemical properties of the pocket residues. Therefore, it is crucial to incorporate the receptor protein information to enable conditional generation. In our architecture, this information is captured by a shared geometric encoder of the protein structure. The encoded information is then fed into the vector field regressor together with the current point information:
% \begin{align}
%     \hat{u}=v_t^M(x_t^M,\mathbf{h}),\mathbf{h}=\mathcal{F}_\text{rec}\left(\left\{a_j,\{\mathbf{x}_{jk}\}_{k=1}^{M}\right\}_{j=1}^{N_\text{rec}}\right)
% \end{align}
% where $M$ are for the four different manifolds with $x^M$ being the point in the manifold.

%% file: Sections/4_experiment.tex
\section{Experiment}

% \input{Tables/1-codesign}
% \input{Tables/2-packing}
% \input{Tables/3-fixbb}

% \paragraph{Overview}

\input{Tables/1-codesign}

\begin{figure*}[!htbp]
    \begin{minipage}{0.4\linewidth}
        \centering     \includegraphics[width=\linewidth]{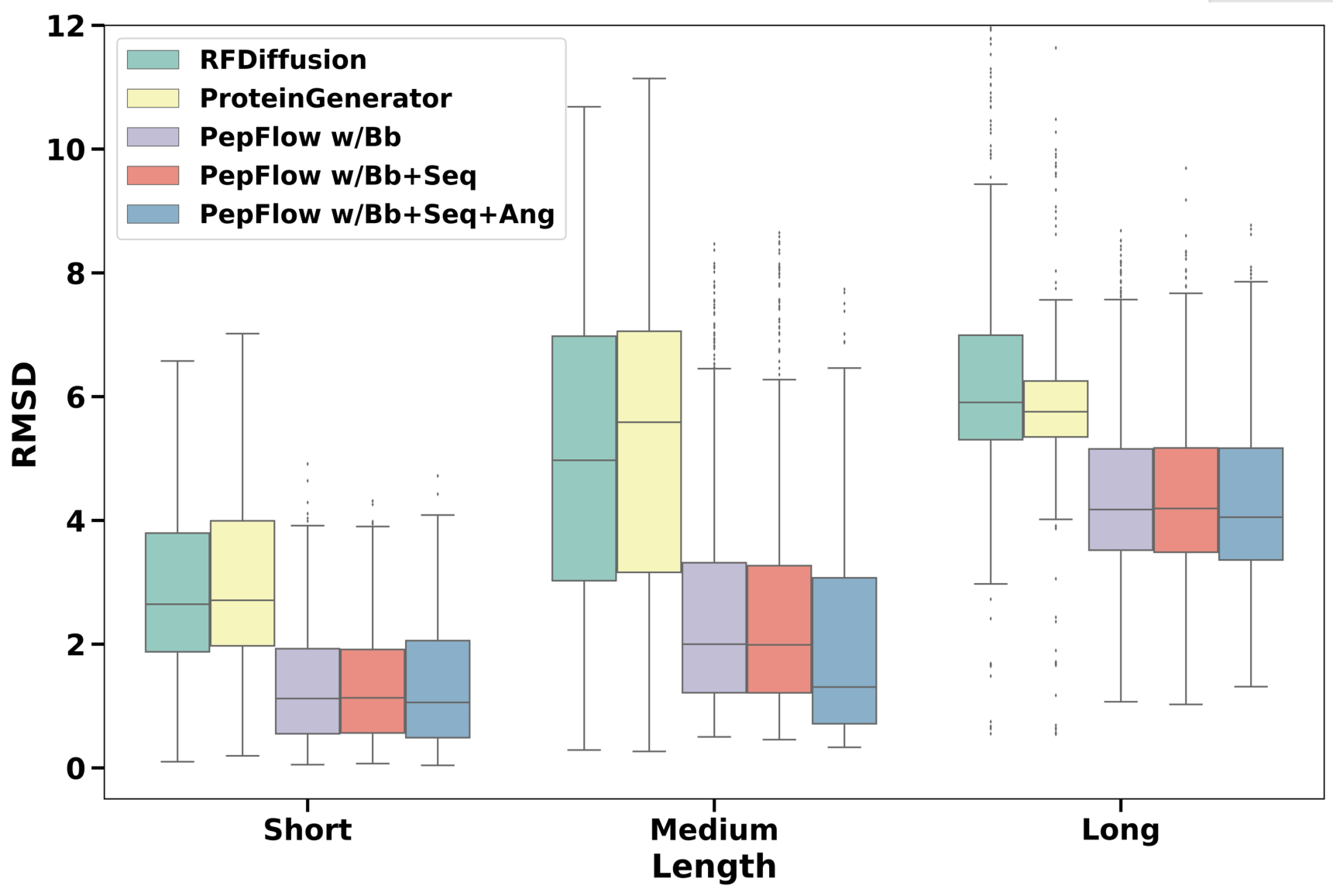}
    \end{minipage}\hfill
    \begin{minipage}{0.27\linewidth}
        \centering
        \includegraphics[width=\linewidth]{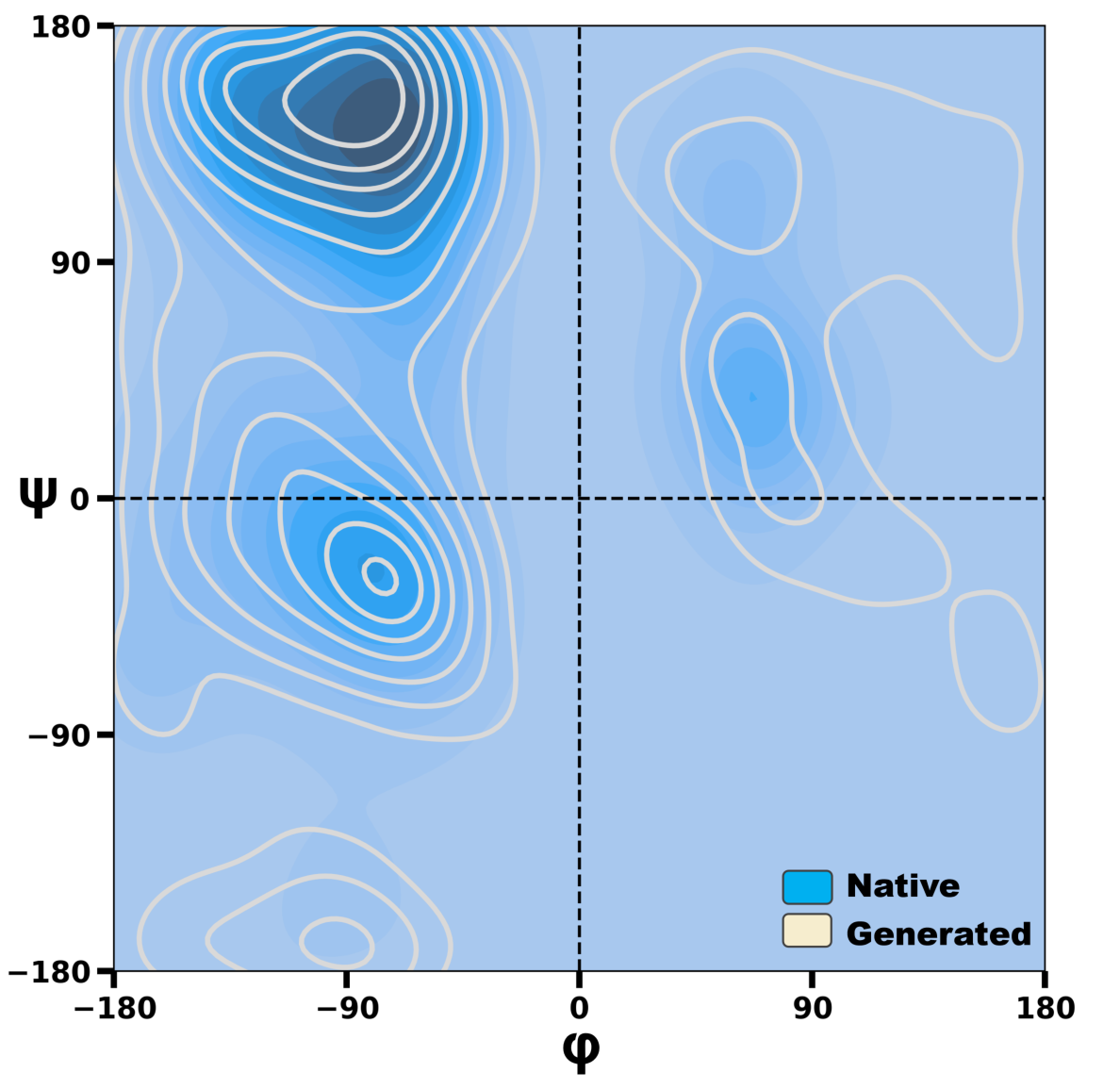}
    \end{minipage}\hfill
        \begin{minipage}{0.33\linewidth}
        \centering
        \includegraphics[width=\linewidth]{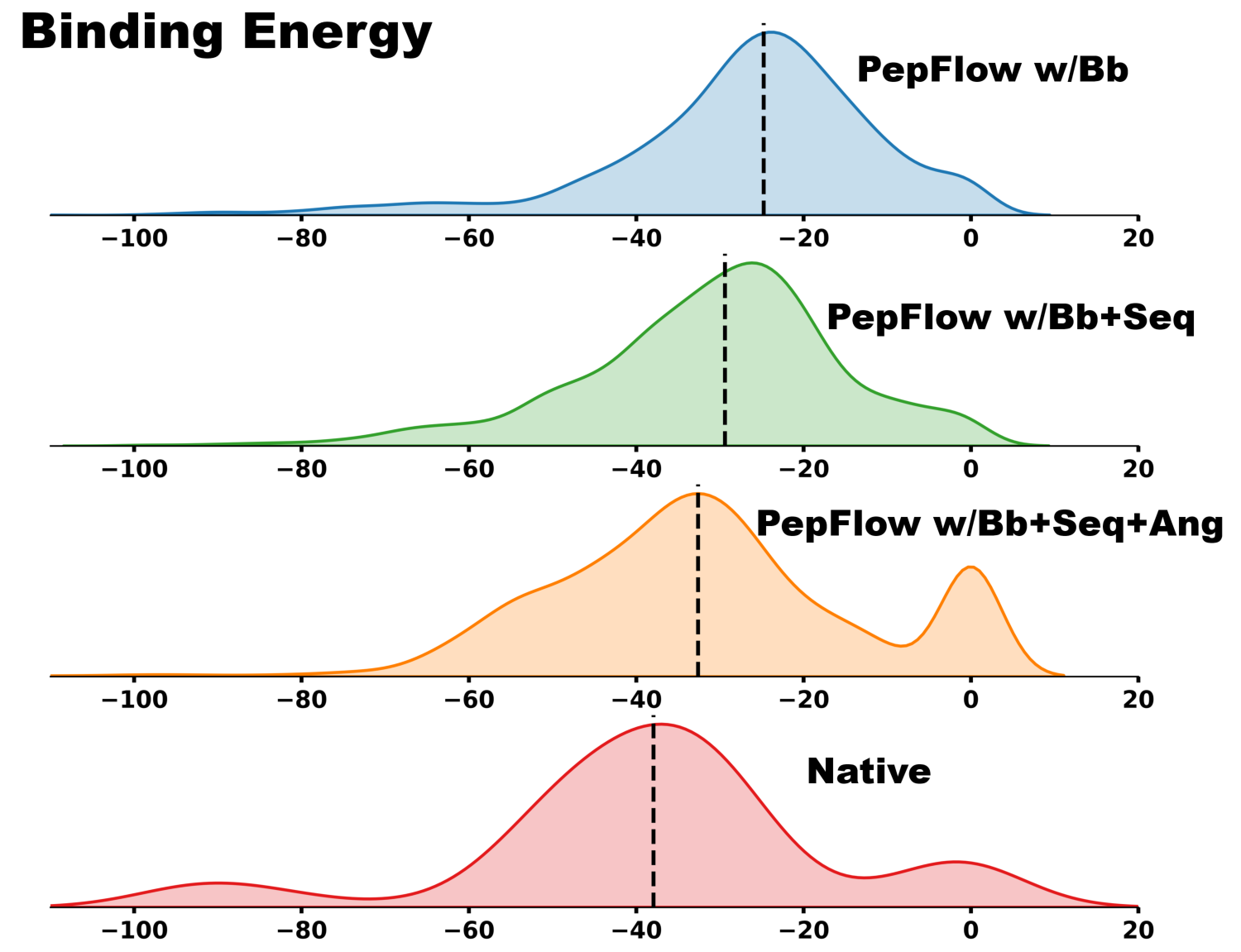}
    \end{minipage}
    \vspace{-1em}
    \caption{\textbf{Left}: RMSD of designed peptides of different lengths. (Short: 3-9, Medium: 10-14, Long: 15-25) \textbf{Middle}: Ramachandran plot of PepFlow generated and native peptides. \textbf{Right}: Binding energy distributions of generated and native peptides. (lower is better)}
    \label{fig:rmsd}
    \vspace{-1em}
\end{figure*}

\begin{figure}[!htbp]
    \centering
\includegraphics[width=0.98\linewidth]{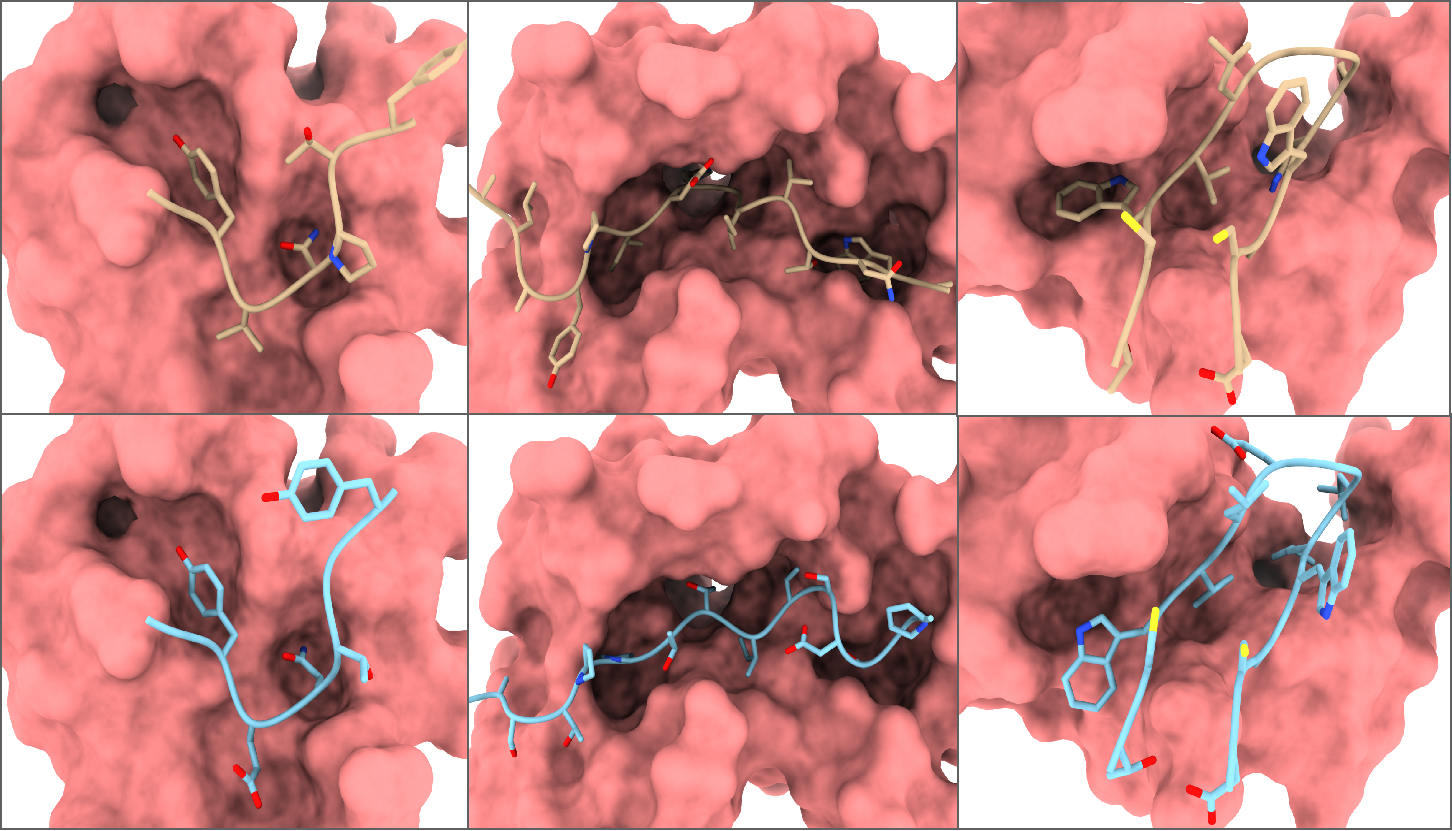}
    \vspace{-1em}
    \caption{Three examples of the generated peptides. \textbf{Top}: native peptides; \textbf{Bottom}: generated peptides. PDB: 3MXY, 6OX4, 5DJY.}
    \label{fig:cases}
    \vspace{-1.em}
\end{figure}

In this section, we conduct a comprehensive evaluation of PepFlow across three tasks: (1) Sequence-Structure Co-design, (2) Fix-Backbone Sequence Design, and (3) Side-Chain Packing. We introduce a new benchmark dataset derived from PepBDB \cite{wen2019pepbdb} and Q-BioLip \cite{wei2024q}. After removing duplicate entries and applying empirical criteria (e.g. resolution $\textless 4\AA$, peptide length between $3-25$), we cluster these complexes according to $40\%$ peptide sequence identity using mmseqs2 \cite{steinegger2017mmseqs2}. This results in $8,365$ non-orphan complexes distributed across $292$ clusters. To construct the test set, We randomly select $10$ clusters containing $158$ complexes, while the remaining complexes are used for training and validation. 

We implement and compare the performance of three variants of our model. \textbf{PepFlow w/Bb} only samples backbones; \textbf{PepFlow w/Bb+Seq} is used for modeling backbones and sequences jointly; and \textbf{PepFlow w/Bb+Seq+Ang} finally model the full-atom distributions of peptides. Experimental details and additional results are provided in Appendix \ref{sup:exp}.

\subsection{Sequence-Structure Co-design}

This task requires the generation of both the sequence and bound-state structure of the peptide based on its target protein. Models take the full-atom structure of the target protein as input and generate bound-state peptides. We generate $64$ peptides for each target protein for every evaluated model.

\textbf{Baselines} We use two state-of-the-art protein design models as baselines. RFDiffusion \cite{watson2022broadly} generates protein backbones, and sequences are later predicted by ProteinMPNN \cite{dauparas2022robust}. ProteinGenerator \cite{lisanza2023joint} improves RFDiffusion by jointly sampling backbones and corresponding sequences. These two methods do not consider side-chain conformations.

\textbf{Metrics} We assess the generated peptides from three perspectives. (1) \textbf{Geometry}. The generated peptides should exhibit sequences and structures similar to those of the native ones. The amino acid recovery rate (\textbf{AAR}) measures the sequence identity between the generated and the ground truth. The root-mean-square deviation (\textbf{RMSD}) aligns the complex by peptide structures and calculates the C$_\alpha$ RMSD. The secondary-structure similarity ratio (\textbf{SSR}) calculates the proportion of shared secondary structures. The binding site ratio (\textbf{BSR}) is the overlapping ratio between the binding site of the generated peptide and the native binding site on the target protein. (2) \textbf{Energy}. We aim to design high-affinity peptide binders that stabilize the protein-peptide complexes. \textbf{Affinity} is the percentage of the designed peptides with higher binding affinities (lower binding energies) than the native peptide, and \textbf{Stability} calculates the percentage of complex structures that are at more stable (lower total energy) states than the native ones. The energy terms are calculated by Rosetta \cite{alford2017rosetta}. (3) \textbf{Design}. \textbf{Designablity} measures the consistency between generated sequences and structures by counting the fractions of sequences that can fold into similar structures as the corresponding generated structure. We utilize ESMFold \cite{lin2023evolutionary} to refold sequences and use C$_\alpha$ RMSD \textless 2\AA as designable criteria. \textbf{Diversity} is the average of one minus the pair-wise TM-Score \cite{zhang2005tm} among the generated peptides, reflecting structural dissimilarities.

\textbf{Results} As indicated in Table \ref{tab:codesign}, PepFlow is capable of generating peptides that closely resemble native ones with improved binding affinities. The distributions of generated backbone torsion angles also closely align with the native peptide distributions (see Figure \ref{fig:rmsd}). Since the generation of structure and sequence is decoupled, RFDiffusion exhibits the lowest recovery, whereas PepFlow achieves lower RMSD and higher AAR by incorporating the sequence modality. Moreover, the explicit modeling of side-chain conformations effectively captures the fine-grained structures in protein-peptide interactions, enhancing the ability of generated peptides to accurately bind to designated binding sites. Consequently, the proportion of peptides with higher affinity is also increased. We also observe that baselines outperformed our models in terms of Stability and Designability, probably because they are trained on the entire PDB structures and are biased toward structures with more stable motifs. Additionally, we note that modeling more modalities leads to a slightly lower diversity of refolded structures. Nevertheless, our results demonstrate that incorporating sequence and side-chain conformation, beyond modeling the backbone, significantly improved the model's performance, underscoring the importance of full-atom modeling.

% \paragraph{Visualizations}
\textbf{Visualizations} We further present three examples of generated full-atom peptides generated by full-atom PepFlow in Figure \ref{fig:cases}. We observe that PepFlow consistently produces peptides with topologically resembling geometries, regardless of the native length. Remarkably, the generated peptides exhibit similar side-chain compositions and conformations, facilitating effective interaction with the target protein at the correct binding site. 

\subsection{Fix-backbone Sequence Design}
This task involves designing peptide sequences based on the structure of the complex without side-chains. We generate $64$ sequences for each peptide using each model. For our models, we apply the partial sampling scheme to recover the sequence. 
% \paragraph{Baselines}

\input{Tables/3-fixbb}

\textbf{Baselines} We use two inverse folding models which can design peptide sequences conditioned on the rest of the complex as our baselines: \textbf{Protein-MPNN} \cite{dauparas2022robust}, a GNN-base model, and \textbf{ESM-IF} \cite{hsu2022learning}, based on GVP-Transformer \cite{jing2020learning}. 

\textbf{Metrics} In addition to \textbf{AAR} used in co-design task, we calculate \textbf{Worst} as the lowest recovery rate. \textbf{Likeness} is the negative log-likelihood score of the sequence from ProtGPT2 \cite{ferruz2022protgpt2}, indicating how closely the generated sequence aligns with the native protein distribution. \textbf{Diversity} represents the average pairwise Hamming Distance between generated sequences.

% \paragraph{Results}
\textbf{Results} As presented in Table \ref{tab:fixbb}, our models achieve higher recovery rates and better diversities, showcasing the effectiveness of our proposed simplex flow matching. We also observe a slight decrease in the recovery rate with the inclusion of angle modeling. This might be attributed to the fact that some amino acids share similar side-chain compositions and physicochemical properties (e.g. Ser and Thr, Arg and Lys), leading the model to generate a physicochemically feasible but different residue. Generating both the sequences and side-chain angles leads to higher likeness and lower diversity.

\subsection{Side-chain Packing}
This task predicts the side-chain angles of the peptide. We generate $64$ side-chain conformations for each peptide by each model and apply the partial sampling scheme of our model to recover side-chain angles.

\input{Tables/2-packing}

\textbf{Baselines} Two energy-based methods: \textbf{RossettaPacker} \cite{leman2020macromolecular}, \textbf{SCWRL4} \cite{krivov2009improved},  and learning-based models: \textbf{DLPacker} \cite{misiura2022dlpacker}, \textbf{AttnPacker} \cite{mcpartlon2022end}, \textbf{DiffPack} \cite{zhang2023diffpack}. 

\textbf{Metrics} We use the Mean Absolute Error (\textbf{MAE}) of predicted four torsion angles. Due to the flexibility of side-chains, We also include the proportion of the \textbf{Correct} predictions that deviate within $20^\circ$ around the ground truth.

\textbf{Results} As shown in Table \ref{tab:packing}, our partially sampling PepFlow outperforms all other baselines across all four side-chain angles. Notably, $\chi_1$ and $\chi_2$ angles are easier to predict than $\chi_3$ and $\chi_4$. Our method also achieves the best correct rates, as our proposed toric flow can precisely capture the reasonable distribution of the side-chain angles and the plausible side-chain dynamics during the interaction between peptides and target proteins.

% \subsection{Flexible Docking}

%% file: Tables/1-codesign.tex
% \begin{table*}[htbp!]
%     \centering
%     \caption{Table1}
%     \label{tab:table1}
%     \resizebox{0.95\linewidth}{!}{
%     \begin{tabular}{lcccccccc}
%         \toprule
%         & RMSD $\uparrow$ & SSR & BSR & Designability & Diversity & AAR & STAB & BIND \\
%         \midrule
%         RFdiffusion & 4.17 & 63.86 & 16.71 & 75.2 & 0.6 & 20.14 & 26.82 & 16.53 \\
%         ProteinGenerator & 4.35 & 59.15 & 14.34 & 70.5 & 0.46 & 18.82 & 23.48 & 13.47 \\
%         Ours w/Bb & 2.3 & 82.2 & 82.17 & 42.43 & 0.36 & 50.46 & 10.04 & 18.1 \\
%         Ours w/Bb+Seq & 2.21 & 85.22 & 85.19 & 54.58 & 0.5 & 53.25 & 19.2 & 19.39 \\
%         Ours w/Bb+Seq+Ang & 2.07 & 83.46 & 85.89 & 61.53 & 0.58 & 52.25 & 18.15 & 21.37 \\
%         \bottomrule
%     \end{tabular}
%     }
% \end{table*}

\begin{table*}[!htbp]
    \centering
    \caption{Evaluation of methods in the sequence-structure co-design task.}
    \label{tab:codesign}
    \resizebox{0.99\linewidth}{!}{
    \begin{tabular}{lcccccccc}
        \toprule
         & \multicolumn{4}{c}{Geometry} & \multicolumn{2}{c}{Energy} & \multicolumn{2}{c}{Design} \\
            \cmidrule(lr){2-5} \cmidrule(lr){6-7} \cmidrule(lr){8-9}
        & AAR $\%\uparrow$ & RMSD $\AA\downarrow$ & SSR $\%\uparrow$ & BSR $\%\uparrow$ & Stability $\%\uparrow$ & Affinity $\%\uparrow$ & Designability $\%\uparrow$ & Diversity $\uparrow$ \\
        \midrule
        RFdiffusion & 40.14 & 4.17 & 63.86 & 26.71 & \textbf{26.82} & 16.53 & \textbf{78.52} & 0.38 \\
        ProteinGenerator & 45.82 & 4.35 & 29.15 & 24.62 & 23.48 & 13.47 & 71.82 & 0.54 \\
        Diffusion & 47.04 & 3.28 & 74.89 & 49.83 & 15.34 & 17.13 & 48.54 & 0.57 \\
        PepFlow w/Bb & 50.46 & 2.30 & 82.17 & 82.17 & 14.04 & 18.10 & 50.03 & \textbf{0.64} \\
        PepFlow w/Bb+Seq & \textbf{53.25} & 2.21 & \textbf{85.22} & 85.19 & 19.20 & 19.39 & 56.04 & 0.50 \\
        PepFlow w/Bb+Seq+Ang & 51.25 & \textbf{2.07} & 83.46 & \textbf{86.89} & 18.15 & \textbf{21.37} & 65.22 & 0.42 \\
        \bottomrule
    \end{tabular}
    }
\end{table*}

%bsr, baselines + 30

%% file: Tables/3-fixbb.tex
% \begin{table}[h]
%     \vspace{-1em}
%     \centering
%     \caption{Evaluations of methods in the fix-backbone sequence design task.}
%     \label{tab:fixbb}
%     \resizebox{0.99\linewidth}{!}{
%     \begin{tabular}{lcccc}
%         \toprule
%         & AAR $\%\uparrow$ & Worst $\%\uparrow$ & Likeness $\downarrow$ & Diversity $\uparrow$ \\
%         \midrule
%         ProteinMPNN & 53.28 & 45.99 & -8.42 & 15.33 \\
%         ESM-IF & 43.51 & 36.18 & -8.39 & 13.76 \\
%         PepFlow w/Bb+Seq & \textbf{56.40} & \textbf{46.59} & -8.26 & \textbf{23.38} \\
%         PepFlow w/Bb+Seq+Ang & 54.32 & 44.48 & \textbf{-8.58} & 20.65 \\
%         \bottomrule
%     \end{tabular}
%     }
%     \vspace{-0.5em}
% \end{table}

\begin{table}[h]
    \vspace{-1em}
    \centering
    \caption{Evaluations of methods in the fix-backbone sequence design task.}
    \label{tab:fixbb}
    \resizebox{0.99\linewidth}{!}{
    \begin{tabular}{lccccc}
        \toprule
        & AAR $\%\uparrow$ & Worst $\%\uparrow$ & Likeness $\downarrow$ & Diversity $\uparrow$ & Designbility $\%\uparrow$ \\
        \midrule
        ProteinMPNN & 53.28 & 45.99 & -8.42 & 15.33 & 60.55 \\
        ESM-IF & 43.51 & 36.18 & -8.39 & 13.76 &  53.76\\
        % PiFold & 45.13 & 34.86 & -8.36 & 14.24 &  52.44\\
        % LM-Design & 55.06 & 42.35 & -8.90 & 16.89 & 61.17 \\
        PepFlow w/Bb+Seq & 56.40 & 46.59 & -8.26 & 23.38 & 59.72 \\
        PepFlow w/Bb+Seq+Ang & 54.32 & 44.48 & -8.58 & 20.65 & 65.48 \\
        \bottomrule
    \end{tabular}
    }
    \vspace{-0.5em}
\end{table}

%% file: Tables/2-packing.tex
% \begin{table}[h]
%     \vspace{-1em}
%     \centering
%     \caption{Evaluation of methods in the side-chain packing task.}
%     \label{tab:packing}
%     \resizebox{0.99\linewidth}{!}{
%     \begin{tabular}{lccccc}
%         \toprule
%         & \multicolumn{4}{c}{MSE $^\circ\downarrow$} \\
%             \cmidrule(lr){2-5}
%         & $\chi_1$ & $\chi_2$& $\chi_3$ & $\chi_4$ & Correct $\%\uparrow$ \\
%         \midrule
%         Rosseta & 38.31 & 43.23 & 53.61 & 71.67 & 57.03 \\
%         SCWRL4 & 30.06 & 40.40 & 49.71 & 53.79 & 60.54 \\
%         DLPacker & 22.44 & 35.65 & 58.53 & 61.70 & 60.91 \\
%         PepFlow w/Bb+Seq+Ang & \textbf{17.38} & \textbf{24.71} & \textbf{33.63} & \textbf{58.49} & \textbf{62.79} \\
%         \bottomrule
%     \end{tabular}
%     }
%     \vspace{-0.5em}
% \end{table}

% new table
\begin{table}[h]
    \vspace{-1em}
    \centering
    \caption{Evaluation of methods in the side-chain packing task.}
    \label{tab:packing}
    \resizebox{0.99\linewidth}{!}{
    \begin{tabular}{lccccc}
        \toprule
        & \multicolumn{4}{c}{MSE $^\circ\downarrow$} \\
            \cmidrule(lr){2-5}
        & $\chi_1$ & $\chi_2$& $\chi_3$ & $\chi_4$ & Correct $\%\uparrow$ \\
        \midrule
        Rosseta & 38.31 & 43.23 & 53.61 & 71.67 & 57.03 \\
        SCWRL4 & 30.06 & 40.40 & 49.71 & 53.79 & 60.54 \\
        DLPacker & 22.44 & 35.65 & 58.53 & 61.70 & 60.91 \\
        AttnPacker & 19.04 & 28.49 & 40.16 & 60.04 & 61.46 \\
        DiffPack & 17.92 & 26.08 & 36.20 & 67.82 & 62.58 \\
        PepFlow w/Bb+Seq+Ang & 17.38 & 24.71 & 33.63 & 58.49 & 62.79 \\
        \bottomrule
    \end{tabular}
    }
    \vspace{-0.5em}
\end{table}

%% file: Sections/5_conclusion.tex
\section{Conclusion}
% In conclusion, we proposed PepFlow, a novel flow-based generative model for peptide structure-sequence co-design. Utilizing the multi-modal flowing matching objectives on Riemannian manifolds, PepFlow can generate the full-atom structure and residue sequence while respecting the intrinsic geometric traits of these different modalities. Experiments have demonstrated the superior performance of PepFlow over the baselines across various tasks and metrics. Nonetheless, we are also aware of the limitations of our model. Compared with other large models like RFDiffusion, PepFlow was trained only on a media-size dataset, which may limit its diversity during generation. Furthermore, in the application of peptide binders, we often expect some desirable properties of the peptide, known as \emph{guided} generation, which PepFlow cannot handle yet.

In this study, we introduced \emph{PepFlow}, a novel flow-based generative model tailored for target-specific full-atom peptide design. PepFlow characterizes each modality of peptide residues into the corresponding manifold and constructs analytical flows and vector fields for each modality. Through multi-modal flow matching objectives, PepFlow excels in generating full-atom peptides, capturing the intrinsic geometric characteristics across different modalities. In our newly designed comprehensive benchmarks, PepFlow has demonstrated promising performance by modeling full-atom joint distributions and exhibits potential applications in protein design beyond peptides, such as antibody and enzyme design. Nevertheless, PepFlow faces limitations in diversity during generation, stemming from the deterministic nature of ODE-based flow. Furthermore, its current incapability for property-guided generation, a common requirement in protein optimization, represents an area for improvement. Despite these considerations, PepFlow stands out as a potent and versatile tool for computational peptide design and analysis. 

\section*{Impact Statement}
PepFlow can contribute to the advancement of computational biology, particularly in the field of protein design, offering a valuable tool for designing therapeutic peptides with potential benefits for human health. While our primary focus is on positive applications, such as drug development and disease treatment, we recognize the importance of considering potential risks associated with any powerful technology. There is a possibility that our algorithm could be misused for designing harmful substances, posing ethical concerns. However, we emphasize the responsible and ethical use of our tool, urging the scientific community and practitioners to employ it judiciously for constructive purposes. Our commitment to ethical practices aims to ensure the positive impact of PepFlow on society.

\section*{Acknowledgements}
This work was supported by the National Key Research and Development Program of China grants 2022YFF1203100. 
Thanks to all my friends for their encouragement, help, and love. 

% \paragraph{Social Impact}

%% file: Sections/6_supplementray.tex
\newpage
\appendix
\onecolumn

Code and data are available at \href{https://github.com/Ced3-han/PepFlowww}{https://github.com/Ced3-han/PepFlowww}.

% The appendix is structured as follows: we begin by outlining the implementation details of PepFlow in \ref{sup:imp}, providing explanations for the manifolds we employ in \ref{sec:manifold}, discussing our reparametrized CFM objectives in \ref{sec:reparam}, considering the geometric aspects of backbone frames and side-chain angles in \ref{sec:geometric}, and presenting network details in \ref{sec:network}. Subsequently, we delve into the experimental details \ref{sup:exp}, covering dataset curation in \ref{sec:data}, and providing details on models and baselines. We also include more sampling results \ref{sec:sample} and potential applications of our methods in \ref{sup:app}.

% We also include more sampling results \ref{sec:sample} and potential applications of our methods in \ref{sup:app}.

% Additional results are included in \ref{sec:res}, and sampling cases are shown in \ref{sup:sample}.
%in \ref{sup:mod,sup:met}

\section{PepFlow Implementations}
\label{sup:imp}

\subsection{Manifold Explanations}\label{sec:manifold}
In this subsection, we further introduce some basic concepts of Riemannian manifolds as well as some important properties of the manifolds used by PepFlow. A Riemannian manifold $\langle\mathcal{M},g\rangle$ is a real, smooth manifold $\mathcal{M}$ equipped with a positive-definite inner product $g$ on the tangent space $T_p\mathcal{M}$ at each point $p$. Let $T\mathcal{M}=\bigcup_{p\in\mathcal{M}}\{p\}\times T_p\mathcal{M}$ be the \emph{tangent bundle} of the manifold, a time-dependent smooth vector field on $\mathcal{M}$ is a mapping $u_t:[0,1]\times\mathcal{M}\to T\mathcal{M}$ where $u_t(p)\in T_p(\mathcal{M})$ for all $p\in\mathcal{M}$.

A \emph{geodesic} is a locally distance-minimizing curve on the manifold. The existence and the uniqueness of the geodesic state that for any point $p\in\mathcal{M}$ and for any tangent vector $u\in T_p(\mathcal{M})$, there exists a unique geodesic $\gamma:[0,1]\to\mathcal{M}$ such that $\gamma(0)=p$ and $\gamma'(0)=u$. The \emph{exponential map} $\mathtt{exp}_p:\mathcal{M}\times T\mathcal{M}\to\mathcal{M}$ is uniquely defined to be $\mathtt{exp}_p(u)=\gamma(1)$. The \emph{logarithm map} $\mathtt{log}:\mathcal{M}\times \mathcal{M}\to T\mathcal{M}$ is defined as the inverse mapping of the exponential map such that $\mathtt{exp}_p(\mathtt{log}_p(q))\equiv q,\forall p,q\in \mathcal{M}$. The exponential map and logarithm map are central in terms of interpolation along the geodesic. As we have mentioned in the main text, the time-dependent flow can be compactly written as
\begin{equation}
    \psi_t(p|p_0,p_1)=\mathtt{exp}_{p_0}(t\mathtt{log}_{p_0}p_1)
\end{equation}
It can be verified that the above formula indeed follows the geodesic between $p_0$ and $p_1$ with linearly decreasing geodesic distances between the interpolated point $p_t$ and the target $p_1$. For general manifolds, closed-form formulae for the exponential and logarithm maps are generally not available. However, the manifolds encountered in this work have well-understood geometries which make it possible for efficient training and sampling using the exponential and logarithm maps.

\paragraph{Euclidean manifold} The tangent space of the Euclidean space $\mathbb{R}^d$ is also $\mathbb{R}^d$, and the canonical inner product for $n$-dimensional vectors can be equipped to form a Riemannian manifold. The exponential map and logarithm map on the Euclidean manifold can be explicitly written as
\begin{align}
    \mathtt{exp}_\mathbf{x}(\mathbf{y})&=\mathbf{x}+\mathbf{y}\\
    \mathtt{log}_\mathbf{x}(\mathbf{u})&=\mathbf{u}-\mathbf{x}
\end{align}
Intuitively, the geodesic between two points in the Euclidean space is exactly the line segment joining the two points, and interpolation at timestep $t$ is given linearly by $\psi_t(\mathbf{x}|\mathbf{x}_0,\mathbf{x}_1)=(1-t)\mathbf{x}_0+t\mathbf{x}_1$. This coincides with the common flow-matching loss for image generation, as already demonstrated in \cite{chen2023riemannian}.

\paragraph{Rotation group $\mathrm{SO}(3)$} The 3D rotation group or the special orthogonal group $\mathrm{SO}(3)$ is a compact 3-dimensional Lie group whose differential is a skew-symmetric matrix in the tangent space $\mathfrak{so}(3)$. The canonical choice of the inner product on the tangent space is the half of the induced Frobenius inner product:
\begin{equation}
    \langle A,B\rangle_{\mathrm{SO}(3)}=\frac{1}{2}\langle A,B\rangle_F=\frac{1}{2}\mathrm{tr}(A^\top B),\forall A,B\in\mathfrak{so}(3)
\end{equation}
This equips $\mathrm{SO}(3)$ with a Riemannian structure. The $\mathrm{SO}(3)$ manifold has a constant Gaussian curvature everywhere and is diffeomorphic to a solid ball with antipolar points identified. The exponential map (from the identity rotation $I$) $\mathtt{exp}:\mathfrak{so}(3)\to\mathrm{SO}(3)$ can be realized as the standard matrix exponentiation:
\begin{equation}
    \mathtt{exp}(A)=\exp(A)=\sum_{k=0}^\infty\frac{A^k}{k!},\forall A\in\mathfrak{so}(3)
\end{equation}
Rodrigues' rotation formula gives an equivalent but more compact form of the exponential map as
\begin{equation}
    \mathtt{exp}(A)=I+\frac{\sin\theta}{\theta}A+\frac{1-\cos\theta}{\theta^2}A^2,\forall A\in\mathfrak{so}(3)
\end{equation}
where $\theta=\|A\|_{\mathrm{SO}(3)}=\frac{1}{2}\|A\|_F$ represents the rotation angle. Similarly, the logarithm map (from the identity rotation $I$) $\mathtt{log}:\mathrm{SO}(3)\to\mathfrak{so}(3)$ can be defined as the matrix logarithm as 
\begin{equation}
    \mathtt{log}(R)=\log R=\sum_{k=1}^\infty(-1)^{k+1}\frac{(R-I)^k}{k}
\end{equation}
or more compactly via
\begin{equation}
    \mathtt{log}(R)=\frac{\theta}{\sin\theta}A,\forall R\in\mathrm{SO}(3)
\end{equation}
where $A=(R-R^\top)/2\in\mathfrak{so}(3)$ and $\theta=\|A\|_{\mathrm{SO}(3)}$ represents the rotation angle. Utilizing the sphere geometry, the geodesic distance between two rotations can be determined as $d(R_1,R_2)=\|\mathtt{log}(R_1^\top R_2)\|_F$, and the interpolation can be calculated as $\mathtt{exp}(tA)$.

\paragraph{Special Euclidean group $\mathrm{SE}(3)$} The special Euclidean group $\mathrm{SE}(3)$ comprises of all rigid transformations of rotation and translation. It can be understood as the semidirect product of $\mathrm{SO}(3)$ and a translation vector in $\mathbb{R}^3$
\begin{equation}
    \mathrm{SE}(3)=\mathrm{SO}(3)\ltimes\mathbb{R}^3
\end{equation}
In practice, the interpolated rotation $R_t$ and translation $\mathbf{x}_t$ at timestep $t$ are fed into a common encoder. Different predictor heads are then applied to the hidden representation to predict the rotation and translation vector field. In this way, each head can leverage the information from both the rotation and translation to learn the joint distribution over $\mathrm{SE}(3)$.

\paragraph{Toric manifold} 
% The toric manifold we used is a \emph{flat torus} that inherits the metric as the quotient space $\mathbb{R}^d/(2\pi\mathbb{Z})^d$. Contrary to the doughnut-shaped manifold in 3D space, a flat torus has zero Gaussian curvature everywhere. The fact that a flat torus is a quotient space of the Euclidean space offers a similar way of defining the exponential and logarithm maps as in the Euclidean space:
% \begin{align}
%     \mathtt{exp}_{a}(u)&=(a+u)\%(2\pi) \\
%     \mathtt{log}_{a}(b)&=\mathrm{atan2}(\sin(b-a),\cos(b-a))
% \end{align}
% Indeed, the flat property indicates that the geodesic on such a torus must be a straight line up to a modulus and that the interpolation must be linear with respect to the interpolants. 

A flat torus in $n$ dimensions is characterized by a metric inherited  from its representation as the quotient, $\mathbb {R} ^{n}/L$, where $L$ denotes a discrete subgroup of $\mathbb {R} ^{n}$ that is isomorphic to $\mathbb {Z} ^{n}$. The tangent space of the torus corresponds to the  Euclidean Space $\mathbb{R}^n$. In our study of torsional flow matching, $L$ represents the cartesian product of $2\pi\mathbb {Z}$ \cite{jantzen2012geodesics}.

Since the flat torus inherits its metric from the quotient manifold, the exponential and logarithm maps can be understood as corresponding to their Euclidean counterparts with modulo of $2\pi$. In Eq.33, the logarithm map is represented as an array of angles  that indicate the direction of the geodesic.
\begin{align}
    \label{eq:toric1}
    \mathtt{exp}_{a}(u)&=(a+u)\%(2\pi) \\
    \mathtt{log}_{a}(b)&=\mathrm{atan2}(\sin(b-a),\cos(b-a))
\end{align}
Then we can derive the simplified vector field on the toric manifold, by definition, the \emph{atan2} function returns an angle in $(-\pi,\pi]$, and we can rewrite the logarithm map, which is locally and linearly connecting the displacement vectors:
\begin{align}
    \text{wrap}(u)&=(u+\pi) \%(2\pi)-\pi \\
     \log_a(b)&=\text{wrap}(b-a) 
\end{align}
Then the interpolation can be written as a locally straight line in the embedded Euclidean space connecting the starting point and end point:
\begin{equation}
    \psi_t=\exp_{\chi_0^j}(t\log_{\chi_0^j}\chi_1^j)=(\chi_0^j+t\text{wrap}(\chi_1^j-\chi_0^j))\%(2\pi) 
\end{equation}
Taking the derivative concerning $t$, we obtain the vector field as:
\begin{equation}
    u_t=\frac{\partial\psi_t}{\partial t}=\text{wrap}(\chi_1^j-\chi_0^j)=\text{wrap}\left(\frac{\chi_1^j-\chi_t^j}{1-t}\right)
\end{equation}

\paragraph{Simplex manifold} A $(d-1)$-simplex $\Delta^{d-1}$ can be used to represent the collect of all $d$-class categorical distributions such that $\Delta^{d-1}=\{\mathbf{p}\in\mathbb{R}^d,\|\mathbf{p}\|_1=1,0\le p_k\le 1,1\le k\le d\}$. In this way, each point $\mathbf{p}\in\Delta^{d-1}$ is a $d$-class categorical distribution on which a flow can be defined. Different from the common Euclidean setting, a simplex manifold is a bounded manifold.  If we construct a flow  directly on the simplex, we may encounter the problems of Points moving outside the boundaries. Note that the vector field on the  boundary points outside the simplex. Point moving away from the manifold. Though this can be prevented by  projecting the predicted vector fields onto the tangent space  $\{\mathbf{u}|\sum_{k=1}^n u_k=0\}$. Therefore, even though directly working on the simplex has the advantage of being simpler to implement, we chose to work on the logit space  which is well-defined over $\mathbb{R}^n$ for unconstraint flow  matching. There is, however, one disadvantage of the projective structure of the logit space. In other words, the mapping between the  simplex and the logit space is not one-to-one, as an equivariant class  of a categorical distribution can be identified with  $\mathbf{s}_1\sim\mathbf{s}_1\Leftrightarrow\exists C\in\mathbb{R}  \text{ s.t. } s_{1,k}=s_{2,k}+C, 1\le k\le n$. We addressed this problem by essentially setting the logit of the last class to 0.

We follow \cite{han-etal-2023-ssd} to assume the logit-normal distribution of the categorical distribution of the amino acid types. In other words, we assume that the logits $s_k=\log(p_k/p_{d}),1\le k\le d-1$ follow some normal distribution in $\mathbb{R}^d$. Such a transform provides a diffeomorphism from the interior of the simplex $(\Delta^{d-1})^\circ$ to $\mathbb{R}^{d-1}$ in which the standard Euclidean flow matching can be applied. It can be demonstrated that such a transform assigns the last class $d$ with a logit of 0. As the boundary of the simplex is excluded in such a logit transformation, we instead use soft one-hot encoding by assigning the ground truth class with a const logit value of $2K>0$ and other logits with $0$ to be consistent with the last class $d$. Note that after applying the softmax function, a logit vector represents the same categorical distribution when adding a constant value to all of the logits. Therefore, such an assignment of the logit values coincides with our definition in Eq.\eqref{eqn:logit}. Similarly, if the last class $d$ is the ground truth class, other logits are set to $-2K$ such that the softmax probabilities remain the same. In this way, the categorical flow matching loss will be symmetric for all classes. The common Euclidean exponential and logarithm map and linear interpolation are applied on $\mathbb{R}^{d-1}$ for learning the vector field. The final amino acid types are sampled from the last categorical distribution. For all cases, we empirically set $K=5$.

\subsection{Reparameterized CFM objectives}\label{sec:reparam}
The Riemannian flow matching paper used the formulation of the object of the conditional vector field as
\begin{equation}
    \mathcal{L}(\theta)=\mathbb{E}_{t,p_1(x_1),p_0(x_0)}\|v_\theta(x_t,t)-u_t(x_t|x_1,x_0)\|_g^2
\end{equation}
where the time-dependent model $v_\theta$ directly learns the vector field at timestep $t$. Following the techniques used in various diffusion models, we can instead predict the target data at timestep $t=1$ and reparameterize the objective as
\begin{equation}
    \mathcal{L}(\theta)=\mathbb{E}_{t,p_1(x_1),p_0(x_0)}\|u_t(x_t|\hat{x}_1,x_0)-u_t(x_t|x_1,x_0)\|_g^2
\end{equation}
where $\hat{x}_1=v_\theta(x_t,t)$. In this formulation, the vector field between the reconstructed data $\hat{x}_1$ and the noise $x_0$ are calculated and compared to the ground truth vector field. Specifically, for flowing matching on Euclidean manifolds, we have $u_t(\mathbf{x}_t|\mathbf{x}_1,\mathbf{x}_0)=\mathbf{x}_1-\mathbf{x}_0$, and
\begin{equation}
    \mathcal{L}_\text{pos}=\mathbb{E}_{t,p_1(\mathbf{x}_1),p_0(\mathbf{x}_0)}\|(\hat{\mathbf{x}}_1-\mathbf{x}_0)-(\mathbf{x}_1-\mathbf{x}_0)\|_2^2=\mathbb{E}_{t,p_1(\mathbf{x}_1),p_0(\mathbf{x}_0)}\|\hat{\mathbf{x}}_1-\mathbf{x}_1\|_2^2
\end{equation}
which coincides with the mean squared error loss. Similarly, the reparameterized loss of the orientation, dihedral angles, and sequence can be formulated as
\begin{align}
    \mathcal{L}_\text{ori}&=\mathbb{E}_{t,p_1(R_1),p_0(R_0)}\left\|\left(\mathtt{log}_{\hat{R}_t}\hat{R}_1-\mathtt{log}_{R_t} R_1\right)/(1-t)\right\|_{\mathrm{SO}(3)}^2\\
    \mathcal{L}_\text{ang}&=\mathbb{E}_{t,p_1(\chi_1),p_0(\chi_0)}\left\|(\hat{\chi}_1-\chi_1)\%2\pi\right\|_2^2\\
    \mathcal{L}_\text{type}&=\mathbb{E}_{t,p_1(\mathbf{s}_1),p_0(\mathbf{s}_0)}\left\|\hat{\mathbf{s}}_1-\mathbf{s}_1\right\|_2^2
\end{align}
Empirically, such parameterization makes the model more numerically stable while $t$ is close to 1 during sampling. Auxiliary losses based on the ground truth data are also feasible to calculate based on the reconstruction, as we will describe in detail in Appendix \ref{sec:aux_loss}. During training, we take the ground truth peptide state $C^{\text{pep}}$ and predicted peptide state $\overline{C}^{\text{pep}}$to calculate vector fields and CFM losses; during sampling, we take vector fields which start from the prior peptide state $C^{\text{pep}}_{\frac{t-1}{N}}$ to the current predicted peptide state $\overline{C}^{\text{pep}}_{\frac{t}{N}}$ for updating the prior state to the next state $C^{\text{pep}}_{\frac{t}{N}}$.

% Similarly, during sampling, the prediction-induced vector field is used to perform Euler steps to update the current state.

We would also like to mention that since different amino acids have varying numbers of side-chain angles, directly calculating torsion loss based on the real amino acid sequence might lead to potential data leakage. This situation could cause the sequence modality to rely on the number of torsions to predict amino acid types instead of learning the underlying biological connections. Therefore, during training, we pragmatically use the predicted amino acid sequence to guide the computation of torsion loss concerning side-chains. For instance, for the $j$-th amino acid, if its true type is LEU but it is predicted as ARG, we use the four side-chain angles of ARG to calculate the angle-related loss, rather than employing the two side-chain angles of LEU. This approach compels the model to treat the two extra side-chain angles from the real type as the default value of 0. In the generation phase, we directly reconstruct side-chain atomic coordinates based on the predicted amino acid type, using the corresponding number of the side-chain angles and the side-chain atomic composition.

\subsection{Geometric Symmetries}\label{sec:geometric}

We model the position $\mathbf{x}_j$ and orientation $R_j$ of each residue as a point in the Riemannian manifolds $\mathbb{R}^3$ and $\mathrm{SO}(3)$ such that the frame $T_j=(\mathbf{x}_j,R_j)$ lies on the Riemannian manifold $\mathrm{SE}(3)$. When extending to the whole peptide with $n$ residues, the set of $n$ frames should lie on a subspace of the Cartesian product of $n$ manifolds, $\mathrm{SE}(3)^n$ with a well-defined $\mathrm{SE}(3)$-invariant metric. To achieve that, we translate the entire protein-peptide complex by subtracting the mean positions of the $n$ residues in the peptide. This ensures that the $n$ positions of the peptide are in the zero center of mass subspace in Euclidean Space and renders the distribution of the generated peptide frames $\mathrm{SE}(3)$-invariant. The defined flows are also $\mathrm{SE}(3)$-invariant, and the vector fields for positions and orientations are $\mathrm{SE}(3)$-equivariant by the use of an equivariant IPA decoder (see Appendix \ref{sec:network}). For the modeling of torsion angles, side chain angles of some residues are $\pi$-rotation-symmetric, which means $\chi \in \mathbb{S}^1$ and the alternative $\chi + \pi \in \mathbb{S}^1$ will result in the same physical structures. We determine the symmetry of the $j$-th residue by the predicted residue type $\overline{a}^j$, followed by evaluating the toric CFM objectives for both the predicted angles and the alternative angles. The minimum values are then used to optimize our networks.

\subsection{Network Details} \label{sec:network}
In this subsection, we describe the network architecture of the target protein encoder and the flow-based decoder in detail.

\paragraph{Encoder} The encoder takes the 3D structural information of the target protein and output node embeddings and node-pair embeddings. For the node (residue) embeddings, we use a mixture of the following features:
\begin{itemize}
    \item Residue type. A learnable embedding is applied.
    \item Atom coordinates. This includes both the backbone and the side-chain atoms.
    \item Backbone dihedrals. Sinusoidal embeddings are applied.
    \item Side-chain angles. Sinusoidal embeddings are applied.
\end{itemize}
These features are encoded using different multi-layer perceptions (MLPs). The concatenated features are transformed by another MLP to form the final node embedding. For the edge (residue-pair) embeddings, we use a mixture of the following features: 
\begin{itemize}
    \item Residue-type pair. A learnable embedding of $20\times 20$ entries is applied.
    \item Relative sequential positions. A learnable embedding of the relative position of the two residues is applied.
    \item Distance between two residues.
    \item Relative orientation between two residues. Inter-residue backbone dihedral angles are calculated to represent the relative orientation and sinusoidal embeddings are applied.
\end{itemize}
Similarly, these features are encoded using different MLPs and are concatenated features and transformed by another MLP to form the final edge embedding. The encoder is $\mathrm{SE}(3)$-invariant, meaning that its output will be the same regardless of any global rigid transformation. Here we set the hidden dimension of residue embedding as $128$, and the hidden dimension of pair embedding hidden dimension as $64$.

\paragraph{Decoder} The flow-based decoder takes the node embeddings and edge embeddings of the receptor proteins, the current interpolated peptide descriptors, and the timestep $t$. It tries to recover the ground truth peptide descriptors at timestep $t=1$ as we are using the reparameterized CFM objectives (see Appendix \ref{sec:reparam}). The overall architecture is based on the Invariant Point Attention (IPA) \cite{jumper2021highly} which takes the above features and backbone frames as input and applies an invariant attention mechanism to capture the interactions between the receptor protein and the current peptide backbone. Additional MLP encoders for the timestep embedding, the residue sequence embeddings, and the dihedral angle embeddings are applied and fused into the IPA output. Separate MLP decoders then try to recover the ground truth descriptors based on the fused information. Note that some residue types may be inferred from the number of residual dihedral angles. To prevent data leakage, we carefully mask out this information for the prediction of the residue types. Here we set the hidden dimension of residue embedding as $128$, and the hidden dimension of pair embedding hidden dimension as $64$, and we use $6$ blocks of IPA.

\subsection{Auxiliary Losses} \label{sec:aux_loss}
The reparameterized flow matching objectives (Appendix \ref{sec:reparam}) make it possible to enforce additional auxiliary loss on the predicted target data at $t=1$. Specifically, the backbone reconstruction loss $\mathcal{L}_\text{bb}$ is proposed to align the prediction with the ground truth backbone:
\begin{equation}
    \mathcal{L}_\text{bb}^j=\sum_{k_\text{bb}}\left\|(\hat{R}^j_1\mathbf{x}_{0,k_\text{bb}}+\hat{\mathbf{x}}^j_1)-\mathbf{x}_{1,k_\text{bb}}\right\|_2^2
\end{equation}
where $\hat{R}^j_1$ and $\hat{\mathbf{x}}^j_1$ are predictions from the model and $\mathbf{x}_{0,k_\text{bb}},\mathbf{x}_{1,k_\text{bb}}$ are the initial backbone atoms coordinates and the ground truth backbone atom coordinates. The backbone reconstruction loss further helps the model to learn the translation and orientation of each residue.

The torsion angles lie on the flat toric manifold, so it is also feasible to enforce a torsion reconstruction loss as
\begin{equation}
    \mathcal{L}_\text{tor}^j=\sum_{k}\left\|(\hat{\chi}_{k,1}^j-\chi_{k,1}^j)\%(2\pi)\right\|_2^2
\end{equation}
where $\hat{\chi}_{k,1}^j$ is the prediction from the model and $\chi_{k,1}^j$ is the ground truth torsion angle. The modulo inside the norm makes sure that the error is in $[-\pi,\pi]$.

\subsection{Training}\label{sec:train}
%400k
We employ three types of models: \textbf{PepFlow w/Bb}, exclusively trained to model the distribution of the peptide backbone; \textbf{PepFlow w/Bb+Seq}, trained to model the joint distribution of backbone and sequence; and \textbf{PepFlow w/Bb+Seq+Ang}, the full-atom version modeling trained on all four modalities of peptide structure and sequence. Despite sharing the same network architecture, each model has a modified training loss. For instance, \textbf{PepFlow w/Bb} includes only position CFM, orientation, and backbone losses, \textbf{PepFlow w/Bb} includes position, orientation, and backbone losses, \textbf{PepFlow w/Bb+Seq+Ang} includes all CFM losses, backbone losses, and torsion losses. We set $\lambda_\text{pos}=0.5,\lambda_\text{ori}=0.5,\lambda_\text{ang}=1.0,\lambda_\text{type}=1.0,\lambda_\text{aux}=0.25$.

All three models are trained on $8$ NVIDIA A100 GPUs using a DDP distributed training scheme for $40$k iterations. We set the learning rate at $5 \times 10^{-4}$ and the batch size at $32$ for each distributed node. To prevent potential gradient clipping issues that may occur during IPA updating, we apply gradient clipping when using the Adam optimizer.

\subsection{Sampling}\label{sec:sample}

We execute the sampling process of our model on a single NVIDIA A100, employing $200$ equal-spaced timesteps for the Euler step update and simultaneously sampling $64$ peptides for each test case. This parallel sampling strategy significantly boosts efficiency, allowing us to generate multiple peptides concurrently.

For scenarios involving partial sampling such as in fix-backbone sequence design, we initialize the prior state with the native peptide backbone structure. During each iteration, we exclusively update the sequence modality while keeping the backbone structure fixed. This approach ensures a focused exploration of the desired modality.

\section{Experimental Details}\label{sup:exp}

\subsection{Dataset Statistics}\label{sec:data}

\begin{figure}[!htbp]
    \centering
\includegraphics[width=0.6\linewidth]{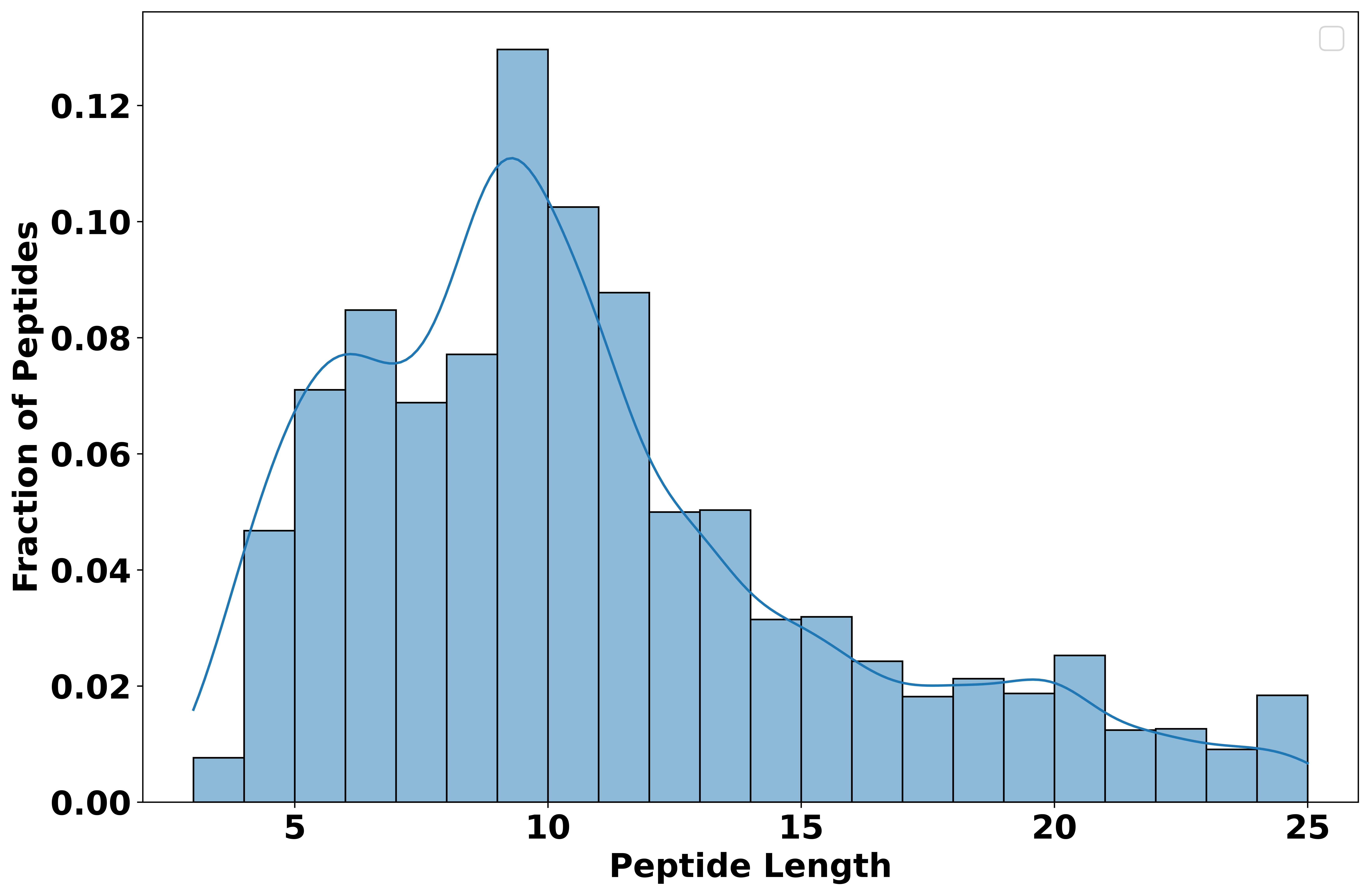}
    \vspace{-1em}
    \caption{Length distribution of the peptide in our dataset.}
    \label{fig:data}
\end{figure}

Our dataset is derived from PpeBDB \cite{wen2019pepbdb} and QBiolLip \cite{wei2024q}. We combine the structures of protein-peptide complexes from these sources and eliminate the duplicate PDBs. Further refinement involves excluding structures with a resolution less than 4\AA, peptides that are too long (exceeding $25$ residues) or too short (less than $3$ residues), and targets with a length not twice as long as that of the peptide. After applying these filters, we reach a set of $10,384$ structures. We then employ clustering based on $40\%$ peptide sequence identity using mmseqs2 \cite{steinegger2017mmseqs2}. This results in $1,557$ clusters. Due to the presence of numerous orphan peptide clusters, we remove clusters with fewer than $5$ items, resulting in a final dataset with $292$ clusters containing $8,365$ complex structures. For efficient and unbiased evaluation, we randomly select $10$ groups from clusters with item numbers ranging from $10$ to $50$ to create our test set of $158$ structures. The remainder of the dataset is allocated for training and validation. Note that for the input of generative models, we derive the binding pocket of the peptide in each complex. The binding pocket is defined as the residues in the target protein which has heavy atoms lying in the $10$\AA radius of any heavy atom in the peptide. The length distribution of peptides is presented in Figure \ref{fig:data}.

\subsection{Methods Details}
\label{sup:mod}

\subsubsection{Sequence-Structure Co-Design}

We remove the peptide in each protein-peptide complex, and the model takes the full-atom target protein structure as input, trying to recover both the sequence and the structure of the corresponding peptide. We generate $64$ peptides for each protein target for every evaluated model. All methods are evaluated on a single NVIDIA A100. Note that the length of the generated peptide is predefined to be the length of each native peptide.

\paragraph{PepFlow} We evaluate all three types of our models. \textbf{PepFlow w/Bb} is used for sampling peptide backbones, and sequences are designed by ProteinMPNN \cite{dauparas2022robust} (1 sequence for each peptide). \textbf{PepFlow w/Bb+Seq} jointly generates backbones and sequences, and \textbf{PepFlow w/Bb+Seq+Ang} is the full-atom version, which designs backbones, sequences, and side-chains.

\paragraph{RFDiffusion}
RFDiffusion \cite{watson2022broadly} utilizes pre-trained weights from RoseTTAFold and undergoes training for generating protein backbone structures through the denoising diffusion process. For designing peptide binders, we employ the official implementation of RFDiffusion. Specifically, we apply $200$ discrete timesteps for the diffusion process, resulting in the generation of $64$ peptides for each target protein. It is important to note that, for the sake of fair comparison, we do not consider the addition of hotspot residues on the target protein during the generation process. Subsequently, the sequences of the generated peptide backbone structures are predicted using ProteinMPNN \cite{dauparas2022robust}.

\paragraph{ProteinGenerator}
ProteinGenerator \cite{lisanza2023joint} represents an updated version of RFDiffusion, capable of concurrently generating protein backbone structures and their corresponding sequences. We utilize the official inference scripts, employing $200$ diffusion timesteps to design $64$ peptides for each target protein. Notably, for a fair comparison, we refrain from incorporating additional hotspot and DSSP information during the generation process.

\paragraph{Diffusion}
We also evaluated a diffusion-based peptide generative model that was trained from scratch using the same settings as our models and possesses similar architectures to the diffab model \cite{luo2022antigen}.

\subsubsection{Fix-Backbone Sequence Design}

We mask the sequence of the peptide in each protein-peptide complex, and models should predict the sequence based on the peptide backbone structure with a full-atom protein target structure and sequence information. We sample $64$ sequences for each peptide using the evaluated methods and report the average of the metrics over all the test peptides. All methods are evaluated on a single NVIDIA A100.

\paragraph{PepFlow} We evaluate both \textbf{PepFlow w/Bb+Seq} and \textbf{PepFlow w/Bb+Seq+Ang}. Both methods take the peptide backbone with the target protein as input, but \textbf{PepFlow w/Bb+Seq} can only sample sequences, while \textbf{PepFlow w/Bb+Seq+Ang} samples both the sequence and the side-chain angles. We do not include \textbf{PepFlow w/Bb} as it can only be used for generating peptide backbones.

\paragraph{ProteinMPNN}
ProteinMPNN is a Graph Neural Network (GNN) based fix-backbone design model \cite{dauparas2022robust}, capable of iteratively predicting protein sequences based on the provided backbone structure. A notable feature of ProteinMPNN is its ability to selectively design specific chains within the complex. When performing Sequence-Structure Co-Design, we utilize the official multi-chain design scripts, setting the sampling temperature to $0.1$. In this context, ProteinMPNN is employed to generate sequences for both RFDiffusion and our \textbf{PepFlow w/Bb} ($10$ sequences for each peptide, but we only select the one with the highest score). Additionally, during fix-backbone design, ProteinMPNN is utilized to design $64$ peptide sequences for each complex structure. 

\paragraph{ESM-IF}
ESM-IF is a fix-backbone design model \cite{hsu2022learning} trained on the CATH dataset and AlphaFold-predicted UniRef 50 structures. For evaluation, we utilize the ESM-IF multi-chain design method to generate $64$ sequences for each complex.

\subsubsection{Side-chain Packing}

We remove the side-chain atoms of the peptide in each protein-peptide complex, and models should predict the side-chain angles of each residue in the peptide. We make predictions $64$ times for each peptide and report the average of the metrics over all the test peptides.

\paragraph{PepFlow} As side-chain packing requires the modeling of side-chain angles, we only evaluate our \textbf{PepFlow Bb+Seq+Ang} by fixing the backbone and sequence to the ground truth and partially sampling for side-chain angles.

\paragraph{RosettaPacker}
We employ the PackRotamersMover in PyRosetta \cite{chaudhury2010pyrosetta}, utilizing the Rosetta energy function \cite{alford2017rosetta} and a rotamer library \cite{shapovalov2011smoothed}. In each test case, we perform packing iterations $64$ times, ensuring a thorough exploration of side-chain conformations.

\paragraph{SCWRL4}
SCWRL4 \cite{krivov2009improved} stands as a widely adopted side-chain packing method, leveraging a backbone-dependent rotamer library within a statistical energy function. For Side-chain Packing evaluation, we download and compile the official program, utilizing it to reconstruct $64$ full-atom structures. 

\paragraph{DLPacker}
DLPacker is a 3D CNN-based model designed for predicting residue side-chain conformations \cite{misiura2022dlpacker}. To reconstruct the side-chains of the peptide, we utilize the official implementation along with the model weights. For each peptide, we perform side-chain packing by generating $64$ structures.

\paragraph{AttnPacker}
AttnPacker incorporates backbone 3D geometry to simultaneously compute all side-chain coordinates using equivariant attention mechanism \cite{mcpartlon2022end}. We adapt the pretrained weights from the official implementation and generate $64$ structures for each complex.

\paragraph{DiffPack}
DiffPack is a diffusion-based generative model which autoregressive pack side-chain angles using diffusion process on toric manifold \cite{zhang2023diffpack}. We adapt the pretrained weights from the official implementation and generate $64$ structures for each complex.

\subsection{Metric Details}
\label{sup:met}

\subsubsection{Sequence-Structure Co-Design}

\paragraph{AAR}
AAR, or Amino Acid Recovery, quantifies the sequence identity between the generated peptide and the native peptide. This metric is computed by evaluating the overlapping ratio of the generated sequence with the native sequence. A higher AAR indicates a closer match between the generated and native peptides in terms of amino acid composition, reflecting the accuracy of the generated sequence.

\paragraph{RMSD}
The Root-Mean-Square Deviation (RMSD) is a standard metric used to compare two protein structures. In our evaluation, the generated peptide within the complex is aligned to the native peptide using the Kabsch Algorithm \cite{kabsch1976solution}. It is important to note that only the peptide structure in the complex is considered for superposition. Subsequently, we calculate the normalized C$_\alpha$ distances between the generated and original peptide, yielding the RMSD value. A lower RMSD indicates a closer structural alignment between the generated and native peptides.

\paragraph{SSR}
As single-chain proteins, peptides cannot form higher-level structures  such as tertiary and quaternary structures; they can only form secondary structures and interact with their receptors through specific motifs \cite{kahn1993peptide,eiriksdottir2010secondary,seebach2006helices}. Secondary Structure Ratio (SSR) assesses the similarity between the secondary structure of the generated peptide and the ground-truth peptide. This metric is computed by determining the ratio of identical entries in the secondary structure labels of the two peptides. The secondary structure labels are obtained using the DSSP software \cite{kabsch1983dictionary}. A higher SSR indicates a closer match in the secondary structure between the generated and native peptides. The use of DSSP for evaluating  secondary structures is inspired by prior works \cite{singh2019peptide,jiang2023explainable}, where DSSP  outputs are treated as ground truth labels of secondary structures. We  will provide additional explanations about this metric.

\paragraph{BSR}
Binding Site Rates (BSR) gauge the interaction similarity between the generated peptide-protein pair and the native peptide-protein pair. Essentially, BSR assesses whether the generated peptide can recognize residues in the target protein similarly to the native peptide, which may imply similar biological functions. We define a residue in the binding site if its C$_\beta$ atom is within a 6\AA radius of any residue in the peptide. BSR is then calculated as the overlapping ratio of the derived binding sites in the generated and native peptides. A higher BSR indicates a closer match in the binding interactions between the generated and native peptide-protein pairs.

\paragraph{STAB}
STAB, or Stability Ratio, is defined as the proportion of designed peptides that exhibit a lower energy score compared to the native complex. The stability of the protein-peptide complex is inversely related to its energy level, as a lower energy score signifies higher stability. Utilizing the \emph{FastRelax} method in PyRosetta \cite{chaudhury2010pyrosetta}, each complex first undergoes relaxation, and the total score is evaluated using the \emph{REF2015} score function. The STAB metric is then determined by calculating the ratio of complexes with the reduced energy scores, highlighting the designed peptides that contribute to the enhanced stability in the protein-peptide complex.

\paragraph{BIND}
BIND represents the percentage of designed peptides exhibiting lower binding energy, indicating a higher binding affinity to the target protein compared to the native peptide. Higher binding affinities are often associated with improved peptide functions. The binding energy is computed using the \emph{InterfaceAnalyzerMover} in PyRosetta \cite{chaudhury2010pyrosetta}, after relaxing the complex and defining the interface between the peptide and target protein. A higher BIND percentage implies that the designed peptides possess enhanced binding affinities, suggesting potential improvements in their functional capabilities.

\paragraph{Designability}
Designability assesses whether a generated peptide structure corresponds to a sequence that can fold into a structure similar to itself. This property is crucial in wet lab experiments where synthesized peptide sequences can be evaluated. We compared the predicted structures from folding models of designed sequences with their native structures. In addition to existing metrics, we have incorporated this assessment to further gauge the performance of different fix-backbone design models. Specifically, we utilized ESMFold to predict the structure of the peptide sequence and subsequently employ Rosetta FlexPep Dock \cite{london2011rosetta,raveh2011rosetta} to dock the predicted structure into the target. The resulting docked structure was then compared with the native peptide structure. We quantified the fraction of generated peptides that exhibit less than 2 $\AA$ RMSD to the native structure as the designability metric.

\paragraph{Diversity}
Diversity is quantified by calculating all the pair-wise TM-scores among the generated peptides for a given target using the original TM-align program \cite{zhang2005tm}. TM-scores represent the structural similarities between peptides. The diversity metric is then defined as 1 minus the average TM-score. A higher diversity value indicates greater structural variation among the generated peptides, showcasing the extent of structural exploration in the design process.

\subsubsection{Fix-Backbone Sequence Design}
\paragraph{Likeness}
Likeness assesses the similarity between the generated sequences and the natural protein sequences. Utilizing a pre-trained protein language model, ProtGPT2 \cite{ferruz2022protgpt2}, each sequence is scored based on the negative log-likelihood. The negative log-likelihood is inversely proportional to the similarity to the natural protein sequence distribution: the lower the negative log-likelihood, the more akin the generated sequence is to the natural protein sequence distribution. This metric offers insights into the fidelity of the generated sequences to the natural protein sequence characteristics.

\paragraph{Diversity}
Diversity is evaluated using the Hamming Distance, which assesses the string distance between two sequences. To quantify diversity among the generated sequences for a given complex, we calculate all the pair-wise Hamming Distances and average them. A higher diversity value indicates a greater dissimilarity among the generated sequences, providing insights into the variety present within the designed peptide set for the target complex.

\paragraph{Designbility}
We compared the predicted structures from folding models of designed sequences with their native structures. In addition to existing metrics, we have incorporated this assessment to further gauge the performance of different fix-backbone design models. Specifically, we utilized ESMFold to predict the structure of the peptide sequence and subsequently employ Rosetta FlexPep Dock to dock the predicted structure into the target. The resulting docked structure was then compared with the native peptide structure. We quantified the fraction of generated peptides that exhibit less than 2 RMSD to the native structure as the designability metric. 

\subsubsection{Side-chain Packing}
\paragraph{MSE}
We use the Mean Squared Error (MSE) to check how well the predicted side-chain angles match the actual values. Since different residues have different numbers of side-chain angles, we calculate MSE separately for each of the four side-chain angles. This helps us understand how accurately the model predicts the angles for different parts of the peptide.

\paragraph{Correct}
Given the flexibility of side-chain angles in space, relying solely on absolute error may not provide a complete picture of the performance of the model. To address this, we define a predicted angle as \emph{correct} if it falls within a $20^\circ$ deviation from its ground truth \cite{zhang2023diffpack}. The metric calculates the proportion of correctly predicted angles for all four side-chain angles, offering a more lenient measure that accounts for the inherent flexibility in the angles.

\subsection{Additional Results}\label{sec:res}

\subsubsection{Sequence-Structure Co-Design}

\begin{figure}[!htbp]
    \centering
\includegraphics[width=0.9\linewidth]{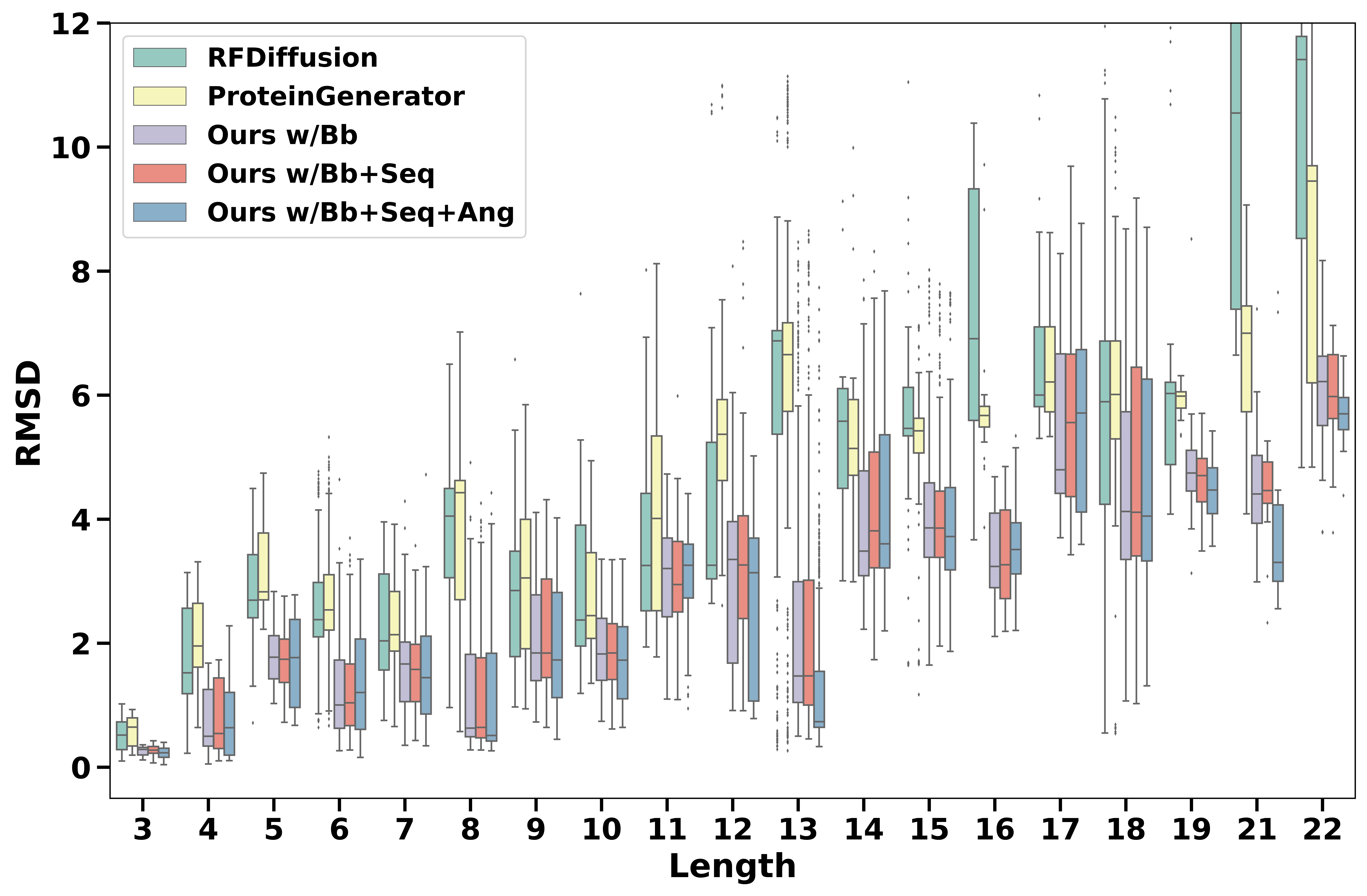}
    \vspace{-1em}
    \caption{RMSD vs Length in sequence-structure co-design tasks.}
    \label{fig:rmsd-len}
\end{figure}

We present the RMSD of the generated peptides for the evaluated methods in Figure \ref{fig:rmsd-len}. We observe that diffusion-based baselines exhibit more variance than our ODE-based method, and all methods perform better in shorter peptide generation than the longer ones.

\begin{figure}[!htbp]
    \centering
\includegraphics[width=0.7\linewidth]{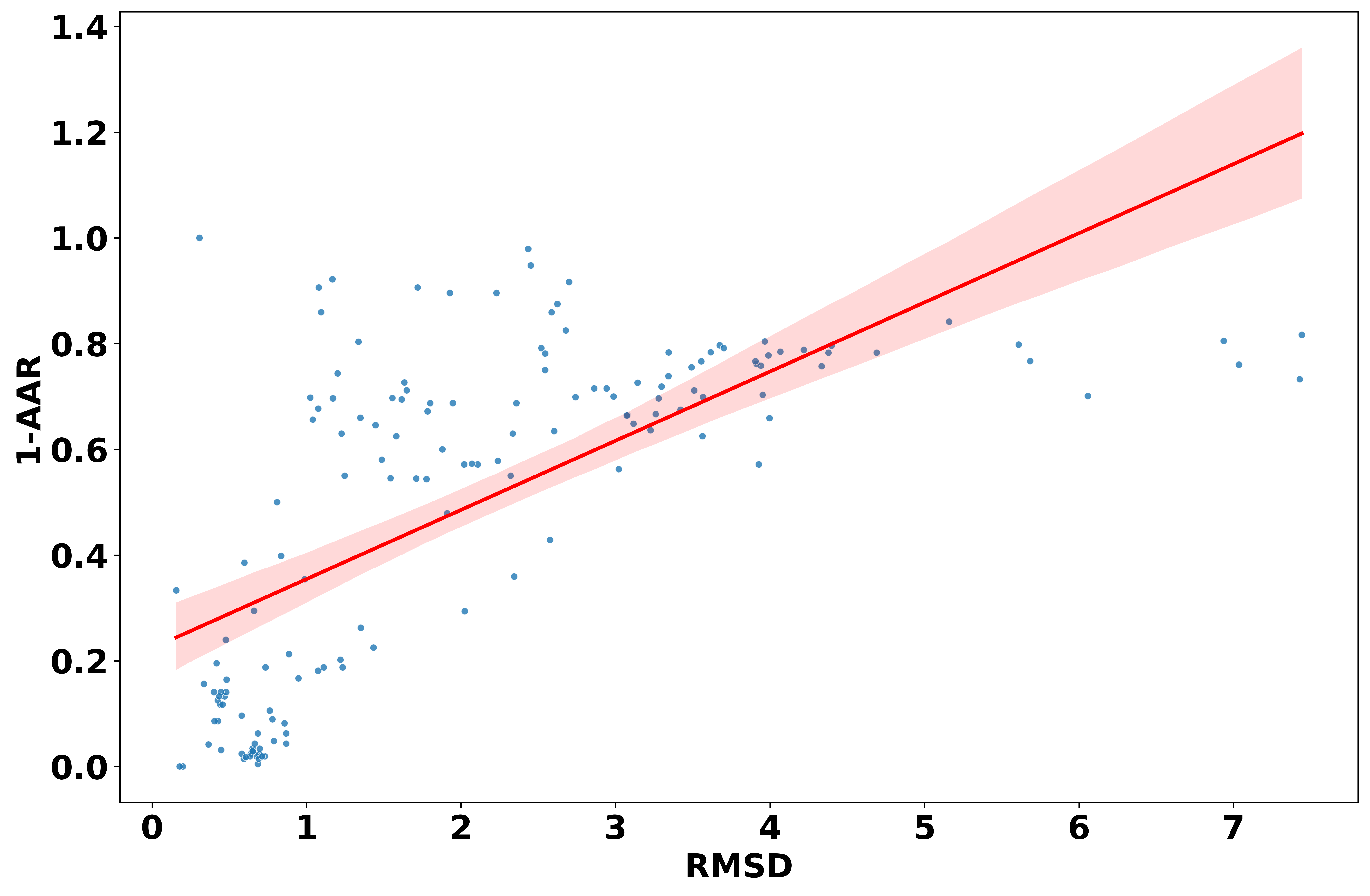}
    \vspace{-1em}
    \caption{Correlation between 1-AAR and RMSD of PepFlow generated peptides, Pearson r is $0.63$. (1-AAR is between 0 and 1).}
    \label{fig:rmsd-aar}
\end{figure}

We also observe a strong correlation between sequence dissimilarity (1-AAR) and RMSD in Figure \ref{fig:rmsd-aar}, with a Pearson correlation coefficient of $0.63$. This underscores the robust consistency between sequence and structure in peptide generation.

\begin{table}[htbp!]
    \centering
    \caption{Additional Results of Sequence-Structure Co-Design}
    \label{tab:add-codesign}
    \resizebox{0.5\linewidth}{!}{
    \begin{tabular}{lcc}
        \toprule
        & C-RMSD $\AA\downarrow$ & TM-Score   \\
        \midrule
        RFdiffusion & 29.02 & 0.31  \\
        ProteinGenerator & 30.88 & 0.33 \\
        PepFlow w/Bb & 6.16  & 0.37\\
        PepFlow w/Bb+Seq & 6.13 & 0.38  \\
        PepFlow w/Bb+Seq+Ang & 5.59 & 0.42 \\
        \bottomrule
    \end{tabular}
    }
\end{table}

Note that the RMSDs reported in our main text and figures are calculated by aligning the generated peptide structures solely to the native peptide. However, given the relatively low Binding Site Ratio (BSR) observed in baseline models, indicating inaccurately placed peptides, we also present the C$_\alpha$ RMSDs by aligning the entire complex (Complex-RMSD \cite{watson2022broadly}), as shown in Table \ref{tab:add-codesign}. Baseline models exhibit subpar performance, suggesting their ability to design peptides but struggle to position them accurately at binding sites. In contrast, our models demonstrate proficiency in generating native-like peptides positioned at similar binding sites to the native peptide, particularly when incorporating side-chain modeling. Additionally, we include the average TM-Score between generated peptide structures, revealing that baselines tend to design more diverse structures, while ours lean towards generating more similar structures. This indicates a trade-off between structure diversities and similarities to the native structure.

\subsubsection{Side-chain Packing}
\label{sup:sc}

\begin{table}[!htbp]
  \centering
  \caption{Mean absolute error of different types of residues.}
  \label{tab:sc-per}
  \begin{minipage}[t]{0.49\linewidth}
    \centering
    \input{Tables/4-1-sc}

  \end{minipage}%
\hfill
  \begin{minipage}[t]{0.49\linewidth}
    \centering
    \input{Tables/4-2-sc}
  \end{minipage}
\end{table}

We also include per residue mean absolute errors of predicted four side-chain angles, as shown in Table \ref{tab:sc-per}. We observe that PepFlow consistently demonstrates better or competitive performance across most cases. Additionally, we notice that for chemically similar amino acids with analogous side-chain compositions, PepFlow exhibits similar performance. This suggests that our designed partial sampling strategy effectively models the distribution of side-chain conformations while adequately considering information from the amino acid sequence modality.

\subsection{More Sampling Results}
\label{sup:sample}

\subsubsection{Sequence-Structure Co-Design}

\begin{figure}[!htbp]
    \centering
    \includegraphics[width=0.9\linewidth]{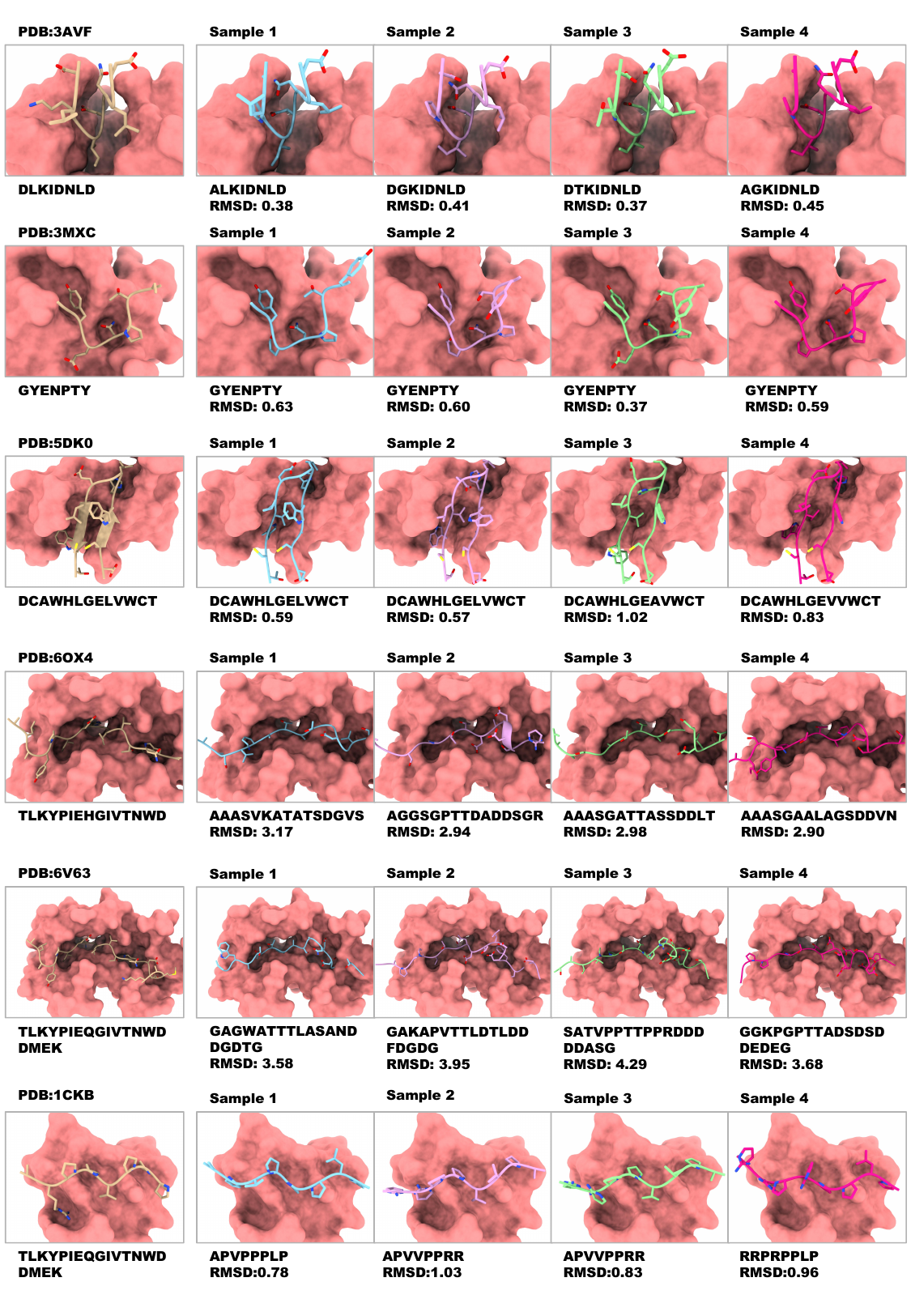}
    \caption{Additional generated peptides by PepFlow.}
    \label{fig:design-case}
\end{figure}

We present additional sampling results of the full-atom \textbf{PepFlow w/Bb+Seq+Ang} for 6 different target protein pockets whose peptide binders have different lengths and topologies, as shown in Figure \ref{fig:design-case}. We observe that PepFlow is capable of designing native-like peptides and using similar residue side-chains to bind key residues in the binding pocket. For short peptides, PepFlow tends to recover the original peptide with low sequence and structure diversities. In contrast, for long linear peptides, PepFlow generates peptides with sequences different from the native, yet they can bind to the same binding site utilizing other residue side-chain groups.

\subsubsection{Side-chain Packing}

\begin{figure}[!htbp]
    \centering
    \includegraphics[width=0.9\linewidth]{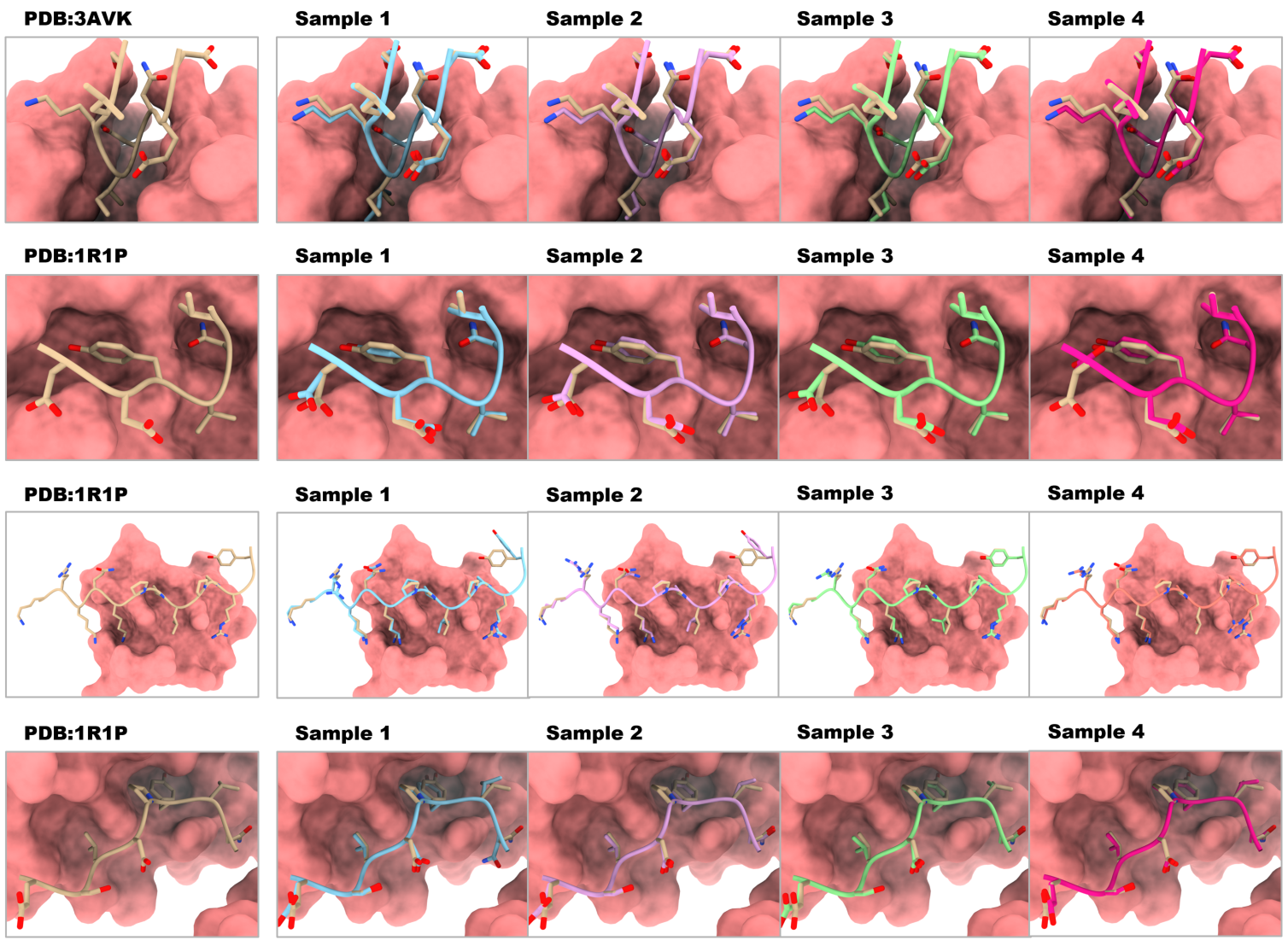}
    \caption{Additional generated side-chain conformations by PepFlow.}
    \label{fig:pack-case}
\end{figure}

We also include several side-chain packing results from \textbf{PepFlow w/Bb+Seq+Ang} in Figure \ref{fig:pack-case}. We observe that PepFlow can accurately recover the side-chains of the residues that are in contact with the target proteins, but performs poorly in modeling the outside residues. This may be attributed to the fact that the contact residues are constrained by binding sites, while the outside ones can exhibit more flexibility.

% \section{Compared to other methods?}

\section{Potential Applications}
\label{sup:app}

Our proposed peptide generation model demonstrates effective applicability to the task of protein binder design, particularly in conjunction with specific receptor proteins. We promise to add more discussion about the application of binder design.

Small protein binders, typically short, single-chain peptides, serve as important candidates in drug discovery and medical applications \cite{muttenthaler2021trends,henninot2018current}. Take, for example, the widely recognized peptide drug GLP-1, which functions by binding to the GLP-1 receptor, aiding in the management of conditions such as diabetes and obesity \cite{cho2011new,skibicka2013central}. Moreover, compared to antibodies, peptides offer advantages in delivery and exhibit lower immunogenicity in the human body. Their utility extends beyond therapeutic purposes; peptide binders can also serve as contrast agents by specifically binding to receptors on certain cells, facilitating precise localization of those cells or tissues within the body \cite{nitin2004functionalization,koudrina2020advances}. The design of these bio-active peptides is fundamental to advancing medical science, and we believe our models can contribute significantly to this endeavor. Nevertheless, challenges persist in these applications, particularly due to the limited availability of complex data, which can impact the generalizability and scalability of the model.

In addition to peptide binders, another class of biomolecules capable of binding to specific proteins is small molecule binders. Designing small molecules based on target structures has been extensively studied within the AI4S community \cite{luo20213d,peng2022pocket2mol}. Notably, considering the atomic interactions between binders and receptors, researchers have identified that small molecules and peptides may utilize similar physical and chemical patterns to interact with key residues in receptors \cite{baines2006peptide,hummel2006translating}. In light of this insight, we posit that developing a unified model capable of designing both small molecules and peptide binders at full-atom resolution represents a promising direction for future research.

While our work primarily focuses on conditional peptide design to evaluate the performance of our proposed full-atom protein design models, it serves as a versatile framework applicable to unconditional protein design and conditional antibody loop design as well, in the full-atom sequence-structure co-design manner. We also acknowledge that the most reliable and accurate evaluation of the generated proteins can only be achieved through wet lab experiments, where in-silico based metrics can be used for selecting good candidates from a large number of generated proteins.

%% file: Tables/4-1-sc.tex
\resizebox{\linewidth}{!}{
\begin{tabular}{ccccccc}
    \toprule
    \textbf{Type} & \textbf{$\chi$} & \textbf{SCWRL4} & \textbf{Rosetta} & \textbf{DLPacker} & \textbf{PepFlow} \\
    \midrule
    LEU & 1 & 16.87 & 26.22 & 25.67 & 30.82 \\
        & 2 & 24.84 & 41.55 & 42.51 & 31.50 \\
    \midrule
    LYS & 1 & 45.31 & 50.43 & 54.45 & 51.40\\
        & 2 & 34.76 & 29.23 & 65.36 & 34.43\\
        & 3 & 20.88 & 30.35 & 47.66 & 31.37 \\
        & 4 & 46.41 & 45.79 & 62.18 & 35.11\\
    \midrule
    MET & 1 & 27.12 & 49.00 & 25.92 & 30.40\\
        & 2 & 62.99 & 60.54 & 33.79 & 44.09\\
        & 3 & 61.00 & 83.63 & 65.07 & 50.01\\
    \midrule
    PHE & 1 & 14.33 & 30.86 & 23.91 & 36.95\\
        & 2 & 98.51 & 110.39 & 43.41 & 23.76\\
    \midrule
    SER & 1 & 62.66 & 48.56 & 43.89 & 35.95\\
    \midrule
    THR & 1 & 36.33 & 53.45 & 30.73 & 22.99\\
    \midrule
    TRP & 1 & 18.89 & 18.64 & 11.58 & 11.21\\
        & 2 & 27.73 & 31.44 & 26.37 & 11.15 \\
    \midrule
    TYR & 1 & 21.04 & 27.99 & 35.46 & 31.32\\
        & 2 & 112.29 & 42.96 & 48.44 & 60.08\\
    \midrule
    VAL & 1 & 26.46 & 25.73 & 20.73 & 18.20\\
    \midrule
    CYS & 1 & 10.16 & 96.40 & 18.45 & 9.49\\
    \bottomrule
  \end{tabular}
}

%% file: Tables/4-2-sc.tex
\resizebox{\linewidth}{!}{
\begin{tabular}{ccccccc}
    \toprule
    \textbf{Type} & \textbf{$\chi$} & \textbf{SCWRL4} & \textbf{Rosetta} & \textbf{DLPacker} & \textbf{PepFlow} \\
    \midrule
    ARG & 1 & 35.50 & 51.48 & 34.67 & 40.80\\
        & 2 & 43.07 & 41.25 & 39.48 & 30.59\\
        & 3 & 61.28 & 54.22 & 48.77 & 32.54\\
        & 4 & 59.14 & 71.71 & 65.48 & 40.16\\
    \midrule
    ASN & 1 & 29.22 & 35.57 & 12.09 & 21.76\\
        & 2 & 33.50 & 33.86 & 31.54 & 23.63\\
    \midrule
    ASP & 1 & 37.75 & 47.06 & 26.56 & 22.12 \\
        & 2 & 76.96 & 78.44 & 84.31 & 40.37 \\
    \midrule
    GLN & 1 & 48.98 & 63.41 & 52.91 & 35.97\\
        & 2 & 56.98 & 66.90 & 66.23 & 38.39\\
        & 3 & 53.08 & 71.95 & 70.21 & 36.01\\
    \midrule
    GLU & 1 & 55.72 & 69.74 & 63.63 & 35.05\\
        & 2 & 34.69 & 32.36 & 30.77 & 41.16\\
        & 3 & 60.28 & 63.67 & 69.88 & 44.47\\
    \midrule
    HIS & 1 & 12.48 & 15.02 & 20.99 & 15.32\\
        & 2 & 30.70 & 37.40 & 28.97 & 51.48\\
    \midrule
    ILE & 1 & 19.77 & 19.84 & 21.53 & 31.55\\
        & 2 & 38.75 & 49.52 & 43.69 & 41.56\\
    \midrule
    PRO & 1 & 12.86 & 12.75 & 10.95 & 16.01\\
        & 2 & 18.97 & 18.30 & 16.61 & 23.55\\
    \bottomrule
\end{tabular}
}